# Chemical capacitor – its concept, functionalities and limits

Łukasz Wolański*[1], Dawid Ciszewski[1], Piotr Szkudlarek[1], José Lorenzana[2] and Wojciech Grochala*[1]

1 Centre of New Technologies University of Warsaw, Żwirki i Wigury 93, 02-089 Warsaw, Poland
2 Institute for Complex Systems (ISC), CNR, 00185 Rome, Italy

**ABSTRACT:** We use density functional theory calculations to study simple but diverse stoichiometries within the novel "chemical capacitor" (CC) setup. We look at main effects occurring in this device, extremes of the physicochemical properties, and we study limits of applicability of this nano-object. In the cases studied, CC permits achieving charge transfer of up to 1.74 e per atom. Tuning of the charge transfer may be achieved via judicious choice of chemical constituents of the CC as well as use of a ferroelectric material as a separator layer. Different classes of chemical systems may be doped, including metallic and nonmetallic elements, and chemical compounds, in certain cases leading to the appearance of superconductivity.

Chemical matter is made up of various combinations of chemical elements, and in very different proportions. However, as Proust noted at the end of the XVIII[th] century, the elements tend to combine in certain proportions that can be expressed with relatively small natural numbers (1). The applicability of this law of definite proportions (also known as the law of constant composition), which is now more than 225 years old, ranges from the smallest systems, such as the $H_2$ molecule, to nanoscopic systems, such as the phosphotungstate anion $[PW_{12}O_{40}]^{3-}$ or an oligoprotein, and mezoscopic systems, such as viral RNA or other supramolecular systems, to macroscopic systems, such as a centimeter-long human chromosomal DNA molecule. To some extent, this important property of chemical matter stems from its largely molecular nature and the fact that atoms tend to form a certain small number of chemical bonds (usually with two electrons per bond). However, most non-molecular chemical systems, such as a crystal of rock salt or calcite, also follow Proust's law. Their behavior confirms the fact that most elements tend to adopt certain integer oxidation states even in extended solids. In fact, these are mostly metals (metallic or semi-metallic elements) that bend the law slightly by forming alloys. Some, such as the coinage metals, can actually mix with each other in any ratio (Table 1). But even in the world of intermetallic compounds, there are many systems that tend to form only certain specific compositions, as is known for various Zintl-Klemm phases (2). These can be regarded as manifestations of Pearson's principle of maximum hardness applied to solids (3).

Apart from metallic alloys, solid solutions are another exception to this law. Solid solutions are quite common among polyelement minerals, and their existence can be easily explained by the Hume-Rothery rules (4). Namely, if two chemical elements have similar atomic sizes, crystal structures, electronegativities and valences, they easily form a solid solution either as elements or as compounds containing their ions. A classic example is lanthanide ores, which can contain all lanthanide metals in one ore. However, the formation of a solid solution does not require strict adherence to the Hume-Rothery rules. For example, the chemical composition of the mineral pyrochlore, which was found in d'Oka (Canada), can be described as follows: $(Na_{2.35}Ca_{11.01}Sr_{0.14}Mn_{0.032}Mg_{0.11}Ce_{1.36}La_{0.27}Nd_{0.34}Y_{0.016}U_{0.064}Th_{0.13}Te_{0.016})(Nb_{11.62}Ta_{0.3}Ti_{2.85}Zr_{0.24}Fe_{0.75})(O_{49.25}(OH)_{0.41}F_{6.34})$, with certainly more constituting elements occurring at much smaller concentrations. The fact that this $Ca(II)_2Nb(V)_2O_7$-type mineral adopts a highly symmetrical, "simple" space group $F$d-3m (No. 227) means that many different cations (and even those with different valences) occupy the same position in the crystal structure. In the case of the pyrochlore in question, monovalent sodium coexists at a Wyckoff position with divalent alkaline earth metals and transition metals, trivalent lanthanides, trivalent and tetravalent actinides and even with the Group 16 semimetal, tellurium" (5). Even non-metals can form solid solutions when combined and quenched at high temperatures, as the example of $C_{1-\delta}B_\delta$ boron-doped diamond shows. The $\delta$ in this formula can vary continuously in a certain wide range between 0 and up to about 0.25 (6). Phosphorus-doped silicon, $SiP_\delta$, is another system in this class. In addition, phases formed by metals and ionic compounds of these metals at elevated temperature, e.g. Bi-metal and $BiCl_3$ or Cs-metal and CsI, can also be quenched (7,8). Non-metals and their ionic salts show a similar tendency to form non-stoichiometric systems, especially for elements with large, soft atoms.

Intermetallic alloys and other solid solutions do not exhaust the list of systems that bend the law of definite proportions (Table 1). At least two other classes of systems of this type are known. One class consists of intrinsically non-stoichiometric compounds. As early as 1866, Graham recognized that hydrogen could dissolve in elemental palladium, leading to a wide range of chemical compositions, $PdH_\delta$ (9). This is a typical feature of the so-called interstitial compounds (here H can fill certain interstices in the Pd network). $Ti_\delta O_\sigma$ and $V_\delta O_\sigma$ (here both $\delta$ and $\sigma$ are non-integer values, near but not necessarily very close to unity) are other examples of non-stoichiometric systems. In fact, the formula of these phases should be written as $(M_{1-\delta}\square_\delta)(O_{1-\sigma}\square_\sigma)$ (M=Ti,V) where $\square$ stands for a vacancy. For $\delta=\sigma$, which is formally "stoichiometric", there are 16% of vacancies of both anions and cations for compounds prepared with different methods (10), i.e. $\delta=\sigma\approx0.16$. These interesting compounds owe its properties to the thermodynamically favorable formation of vacancies at the crystallo-graphic sites of M and O driven by strong electronic correlations (11).

Another family of non-stoichiometric compounds is represented by "intentionally doped" systems. These are mainly human, i.e. artificial, factories. This class is represented by the following stoichiometries, among others: $Li_\delta C$ (12), $Li_\delta MO_2$ M=Mn, Co (13), $La_{1-\delta}Ba_\delta CuO_4$ (14), $Li_\delta HfNCl$ (15), $H_\delta WO_3$ (16), $La_{1-\delta}Ca_\delta MnO_3$ (17), $Ca_{1-\delta}Y_\delta F_{2+\delta}$ (18), $(poly-C_2H_2)(I_3)_\delta$ (19) and $LaH_{10+\delta}$ (20,21). The boundaries between these de facto informal categories are not strict, and $SiP_\delta$ can be considered as a representative of the "intentionally doped" category and at the same time as a representative of the "solid solutions of nonmetals" category.

**Table 1. Examples of chemical systems which break the law of definite proportions.**

|  | Examples |
|---|---|
| Intermetallic alloys | $Cu_{0.03}Ag_{0.25}Au_{0.72}$ |
| Quenched solid solutions of nonmetals | $C_{1-\delta}B_\delta$, $SiP_\delta$ |
| Quenched solid solutions of nonmetals and their salts, or metals and their salts | $Cs_{1+\delta}I$, $CsI_{1+\delta}$, $Bi_{1+\delta}Cl_3$ |
| Solid solutions (compounds) | $(La_{0.15}Ce_{0.25}Pr_{0.13}Nd_{0.47})_2O_3$ |
| Intrinsically non-stoichiometric and/or interstitial compounds | $PdH_\delta$, $Ti_\delta O_\sigma$, $V_\delta O_\sigma$ |
| Deliberately doped compounds | $Li_\delta C$, $Li_\delta HfNCl$, $Li_\delta MO_2$ (M=Mn, Co), $La_{1-\delta}Ba_\delta CuO_4$, $H_\delta WO_3$, $La_{1-\delta}Ca_\delta MnO_3$, $Ca_{1-\delta}Y_\delta F_{2+\delta}$, $(C_2H_2)(I_3)_\delta$, $LaH_{10+\delta}$ |

**Properties of non-stoichiometric compounds.** All the non-stoichiometric compounds mentioned above and others belonging to this family have a number of interesting physicochemical properties. One rather trivial property is density, which can vary widely across the compositional range. Occasionally, as with $Ti_\delta O_\sigma$ where $\delta=\sigma$, the density behaves quite peculiarly, being very different for TiO and $Ti_{0.9}O_{0.9}$, whereas both phases contain Ti and O in the same ratio (1:1). Another reason is the possibility of producing systems with different mechanical properties leading to different uses (just think of alloys and the copper, bronze and iron ages). However, the key feature of all non-stoichiometric materials for modern technologies is that their electrical and sometimes magnetic properties can depend dramatically on δ. Take $C_{1-\delta}B_\delta$: Boron-doped diamond is a black compound with a metallic lustre that is used as a chemically inert electrode material. On the other hand, $SiP_\delta$ or $SiB_\delta$ are usually produced with quite small δ of up to 0.01, but the precise control of δ is crucial for controlling wealth. Imagine our civilization without these n- or p-doped semiconductors and all the devices resulting from the fabrication of n-p junctions, NPJs (central processing units, CPUs, solar cells, diodes, transistors, charge-coupled devices, CCDs, in our cameras, etc.). All other materials mentioned here show the same dependence of key properties on the degree of doping. For example, $Cs_{1+\delta}I$ and $CsI_{1+\delta}$ are both metallic, while the undiluted CsI is an ionic insulator for electricity. $(La_{0.15}Ce_{0.25}Pr_{0.13}Nd_{0.47})_2O_3$ and other salts of the same type (e.g. transition metal compounds) form so-called highly entropy alloys, which exhibit a number of useful properties (22,23). $PdH_\delta$ changes its electrical conductivity so much with δ that Pd can be used to detect traces (ppm) of $H_2$ gas. Other systems are no less important. $Li_\delta C$ is a member of the rich family of intercalated graphites. $Li_\delta CoO_2$ is a classic lithium storage material (first generation) for lithium batteries, $La_{1-\delta}Ba_\delta CuO_4$ is the first known high-temperature superconductor, SC, $Li_\delta HfNCl$ is also a SC, $H_\delta WO_3$ and $Na_\delta WO_3$ are the famous tungsten bronzes, $La_{1-\delta}Ca_\delta MnO_3$ shows huge magnetoresistance, $Ca_{1-\delta}Y_\delta F_{2+\delta}$ is an excellent fluoride ion conductor (solid electrolyte), $(poly-C_2H_2)(I_3)_\delta$ is a prototypical organic conductive polymer, while $LaH_{10+\delta}$ at a very high external pressure of 188 GPa is the first known SC with a critical temperature of 260 K (or –13 ºC), approaching the melting point of ice.

It is obvious that all non-stoichiometric systems have fascinating properties and are of immense practical importance, apart from their key role in basic research. One reason why they exhibit such peculiar properties – although they represent a minority of all chemical substances – is the fact that the formal oxidation state of at least one element in their composition is not fixed to an integer value, but depends continuously on δ. This makes $Li_\delta CoO_2$, for example, very different from $Fe_3O_4$, as the formal oxidation state of the transition metal changes continuously in the former, while it is fixed at 2+ and 3+ in the latter. In other words, non-stoichiometric $Li_\delta CoO_2$ represents intermediate valence (IV) systems, whereas magnetite at ambient conditions (p,T) is a classical "frozen" or mixed valence (MV) (24).

**Preparation of doped materials.** Non-stoichiometric doped materials can be produced using a range of physical and chemical methods. The classical approaches include: (i) co-melting, sometimes in conjunction with quenching; (ii) chemical intercalation (especially of alkali, alkaline earth and lanthanide metals) and electrochemical intercalation (especially of Li or H), (iii) introduction of defects e.g. by irradiation with vis or UV light or high-energy rays (gamma rays), (iv) reaction of material with reactive elements, e.g. $H_2$ ($PdH_\delta$), H *in statu nascendi* ($H_\delta WO_3$), $I_2$ ($(poly-C_2H_2)(I_3)_\delta$), $O_2$ ($YBa_2Cu_3O_{6.5+\delta}$), $F_2$ ($Sr_2CuO_4F_\delta$) etc., and (v) the use of extreme pressures for the synthesis ($LaH_{10+\delta}$). The use of these high-energy methods points to the difficulties that many materials have with doping, as well as their often-variable nature. Another, physiological, method is to place a thin, prefabricated layer of material (sometimes only a single layer) in a strong electric field, e.g. with a field effect transistor (FET) or coming from ionic liquid transistor setup. Such devices can be used to inject electrons or holes into various substances, sometimes resulting in superconductivity (25,26,27).

**Chemical capacitor.** Recently, another doping method has been proposed (28). This method is based on the placement of a (single or multiple) layer(s) of an oxidising agent, OX, near a similar layer(s) of a reducing agent, RED, using an inert separator, SEP, with controllable thickness (Figure 1). It turns out that when the energy of the upper valence band of RED, $E_V^{max}$, is above the energy of the lower conduction band of OX, $E_C^{min}$, a (non-integer) subset of electrons is transferred from RED to OX (Figure 2):

RED + OX → RED$^{\delta+}$ + OX$^{\delta-}$ (Eq.1)

Neglecting bandwidth effects, the energy gain due to this charge transfer is proportional to $E_{CT}= -(E_V^{max} - E_C^{min})\delta$. However, if RED and OX are spatially separated and the charge in uniformly distributed in the RED and OX layer, such a device can be seen as a charged capacitor. A macroscopic electric field $E \propto \delta$ appears between RED and OX which is independent of the distance d among layers. Since the electrostatic energy density is proportional to $E^2 \propto \delta^2$ the energy cost per unit surface to separate the charges is $E_C \propto d\delta^2$. Minimizing $E_{CT}+E_C$, one obtains:

$$\delta \propto 1/d \propto 1/(n+1) \quad (Eq.\ 2)$$

with n the number of SEP layers (Fig. 2, right), thus δ is a monotonic function of the thickness of the separator. Obviously, δ decays to zero for a very thick separator (28).

A device consisting of at least RED, SEP and OX, labeled here as **RED | SEP | OX** (possibly together with a suitable substrate for OX and a top cap for RED), is called a "chemical capacitor", CC (29). This name stems from the fact that the driving force for charge transfer is the difference in chemical potential, μ, between OX and RED (hence "chemical"), while the hyperbolic dependence of δ on the distance between the two is similar to that of the charge, Δq, accumulating in a physical capacitor, PC, as a function of applied voltage (hence "capacitor"). CC is thus a member of a triad, which also consists of the aforementioned PC and the electrochemical cell ECC (Figure 3). While PC converts the difference in electrical potentials, ΔV, to Δq, ECC converts Δμ to ΔV, while CC uses Δμ to generate Δq. In a way, ECC may be thought of as composed of both the CC and the PC in terms of redox-active substrates and metallic electrodes (Figure 3).

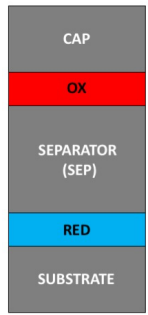

Figure 1. General construct of a chemical capacitor with three indispensable elements: a layer of Oxidizer, a Separator, and a layer of Reducing agent, with optional Substrate and a Cap (which may be constructed from the same or different materials as a Separator).

Of these three, both ECC and PC can in principle be macroscopic, microscopic or nanoscopic components. However, CC is only functional if it is nanoscopic; otherwise, as already said, Δq goes to zero and the main advantages (charge transfer, CT leading to metallization of both OX and RED) disappear.

Note that the method of doping with CC is most similar to that with FET, since the doping is not achieved by a change in chemical composition but by charge injection. I.e. the law of definite proportions is followed, while the properties change due to the transferred charge. However, CC also differs from an FET in that no external electric field is required to create the effect. In addition, two layers (OX and RED) are mutually doped and not separated, which is achieved in two different FETs (independently for OX and RED). And last but not least, the doping disappears when the electric field is switched off in one FET, while it is permanently present in a CC.

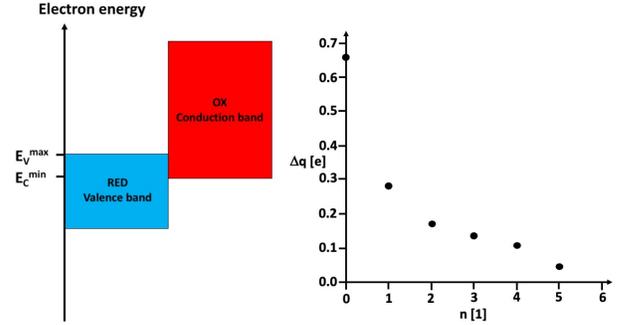

Figure 2. (left) The valence band of RED with its maximum $E_V^{max}$, and the conduction band of OX, with its minimum $E_C^{min}$ falling below $E_V^{max}$; (right) the fractional charge transferred between OX and RED, Δq, depending on the number of separator layers, n, between them (26). Calling c the interlayer spacing and neglecting non uniform effects, the distance between layers is d=(n+1)c.

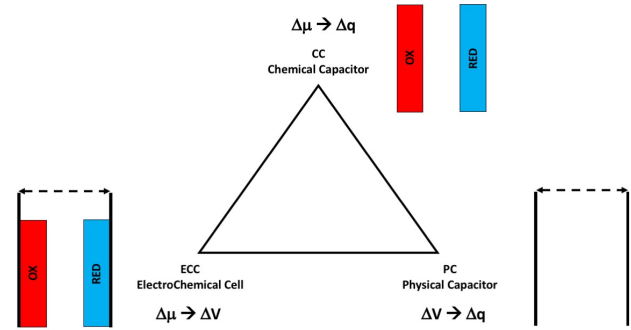

Figure 3. A triad of related devices, ECC, PC and CC. Thick black lines stand for metallic electrodes.

**CCs studied so far by theory.** Only a relatively small number of OX/RED pairs have been theoretically investigated so far. The first systems investigated (28) (Eq.3 and 4) originated from our long-standing goal of metallizing AgF$_2$ and transforming it into an SC (30) (Figure 4):

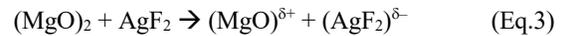
(MgO)$_2$ + AgF$_2$ → (MgO)$^{\delta+}$ + (AgF$_2$)$^{\delta-}$ (Eq.3)
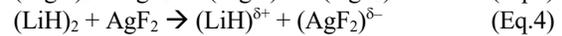
(LiH)$_2$ + AgF$_2$ → (LiH)$^{\delta+}$ + (AgF$_2$)$^{\delta-}$ (Eq.4)

The recent density functional theory (DFT) study (26) has shown the general principle of developing a CC using RbMgF$_3$ perovskite as a suitable SEP material. This insulating fluoride contains metals in their highest available oxidation states and is not oxidized in contact with a strong AgF$_2$ OX. On the other hand, it is also not easily reduced and even resists direct contact with LiH, a strong RED. It also has a large band gap (which translates to a relatively small static dielectric constant

(ε)), so it can withstand a huge electric field that occurs during charge transfer relatively easily. At the same time, the respective lattice constants are very similar to each other, which offers the possibility of producing such systems by epitaxy. Note that a good match of lattice constants is almost always necessary to perform quantum mechanical calculations with periodic boundary conditions, i.e. with programs suitable for extended solids (1-, 2- or 3-dimensional).

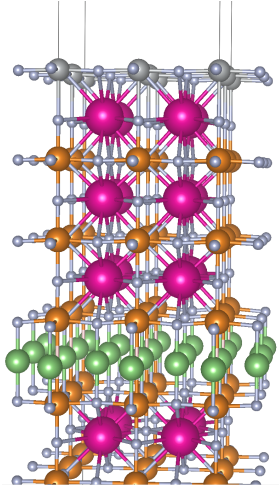

Figure 4. The first CC system studied theoretically, consisting of OX $AgF_2$, RED LiH, and SEP and substrate $RbMgF_3$ (28). Ag – large gray, F – small gray, Rb – large violet, Mg – large orange, Li – large green, H – small white balls.

The original study (28) showed how to manipulate Δq by not only varying the thickness of the SEP layer, but also by carefully selecting the RED component while keeping OX constant. It is noteworthy that the weak reducing agent MgO results in a smaller Δq than the stronger reducing agent LiH with the same distance between the OX and RED monolayers. Of course, one could alternatively change OX while keeping RED constant, which would have a similar effect.

Equation (2) for δ or, more precisely, for the charge Δq per metal atom, can be improved taking into account bandwidth and interaction effects through the self-capacitance per metal atom, $C_V$ ($C_C$) of the MgO ($AgF_2$) layer and a geometrical capacitance $C_M=(\varepsilon a^2)/4\pi d$ with a being the Ag-Ag intra-sheet plane distance and ε being the static dielectric constant. These yield Eq.5 (28):

$$\Delta q = (E_V^{max} - E_C^{min})/[e^2(1/C_V + 1/C_C + 1/C_M)] \quad (Eq.5)$$

where e is electron charge, while $E_V^{max}$ and $E_C^{min}$ were defined above. The above-mentioned changes of redox-active materials obviously alter the $E_V^{max}$–$E_C^{min}$ term.

The subsequent theoretical study (31) focused on other ways of doping $AgF_2$, including electron and hole doping. Since $AgF_2$ is a very strong OX, this study showed that doping $AgF_2$ with holes is practically impossible, while doping with electrons is indeed easy.

The most recent study (32) has been centered on the possibility of metallizing ionic hydrides, such as the LiH mentioned above, or the related NaH and $MgH_2$, to achieve SC. *Ca.* twenty different systems were investigated and their $T_C$ values calculated. Since these calculations are very computationally intensive, the authors were forced to restrict the entire CC structure to the RED layer surrounded by a small number of SEP layers (Figure 5). Note that the calculations for the entire CC setup such as one shown in the Figure 1, would be technically impossible as they are overly CPU-demanding. In this case, the level of hole doping (charge) represented a variable, and the Jellium model was used for charge equalisation within the unit cell. The study showed that it is possible to achieve an SC of up to 17.5 K in hole-doped LiH without external pressure. In addition, certain systems were found to be resistant to gigantic hole doping of up to 0.72 $h^+$ per H atom, which in itself is a fascinating result.

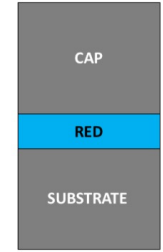

Figure 5. A truncated CC used in (32) to study metallization of ionic hydrides (RED) and employing a jellium model; positive charge has been doped gradually to the RED layer in this study without the necessity to involve OX nor a variable-thickness SEP.

**Experimental realization of a CC.** When the concept of a CC was originally put forward, CC seemed to be a purely theoretical object. However, as the manuscript (28) was in press, we learned of an experimental study from 2015 (33). These authors succeeded in perforating graphene using molecular $O_2$ chemisorbed on the surface of iridium metal; silicon dioxide served as the SEP (Figure 4). The work (33) provides a clear experimental demonstration of the CC effect, as increasing thickness of silica leads to a decrease in hole doping of graphene. The effects are three orders of magnitude smaller than those calculated for $AgF_2$/LiH monolayers in (28). However, it is important to realize that graphene and $O_2$ form a pair with a relatively small difference in chemical potentials; on the contrary, $AgF_2$ is among the strongest OXs, while LiH is a powerful RED substance. In addition, silica SEP cannot be deposited layer by layer, and a very large thickness of SEP of tens of nanometers was used in (33). On the other hand, the large Δq calculated in (28) was observed at a distance of less than one nanometer.

Although the charge transfer in the experimental study (33) is quite small, it brings another aspect that is missing in (28). Namely, the experimentally studied iridium/$O_2$/silica/graphene system is **not** epitaxial, as both the lattice constants and the crystallographic systems are drastically different for all components. And yet CT is still observed. This proves that the requirement of epitaxy (28) is a purely technical limitation related to the theoretical tools used in this work (i.e. the need for periodic boundary conditions), but it is not an imperative in experimental studies.

A related realization is the appearance of a two-dimensional electron gas at the interface of different oxides, like $LaAlO_3$/$SrTiO_3$ driven by a polar catastrophe (34). In this case the $LaAlO_3$ is grown layer by layer on top of a flat $SrTiO_3$ substrate alternating $(LaO)^+$ layers with $(AlO_2)^-$ layers. This can be seen as an RED-OX-RED-OX… multilayer system. If the number of RED and OX layers are equal, the bulk RED and OX neutralize and effectively act as a separator. On the other hand, in an ionic picture with formal valences, the first RED layer remains positively charged and the last OX layer negatively charged. Thus, we have again an effective capacitor with energy growing as d, the distance between the first RED and the last OX layer. For small d this charged capacitor is stable because LaAlO$_3$ behaves as an ionic insulator. However, for d large enough (more than 3-unit cells) (35) it is convenient to pay the band gap energy and transfer ½ an electron per surface metal atom from the OX to the RED layer region and save the electrostatic energy (35,36). The transferred charge actually dopes the SrTiO$_3$ substrate, creating a high mobility electron gas. Because the starting system at small d is an insulator, contrary to the CC of Ref. (28), doping appears increasing the distance between charged layers. It is believed that oxygen vacancies also play an important role in creating the two-dimensional metallic state in LaAlO$_3$/SrTiO$_3$ interfaces (36).

Finally, another very interesting and related development are Janus two-dimensional materials. In this case, OX and RED molecules are placed at the opposite side of a two-dimensional material like graphene. These interesting structures are predicted to have strong piezoelectric properties and strong spin-orbit effects (due to the internal electric field) which are useful for device applications (37). Such effects are common to all kinds of CC due to the absence of inversion symmetry, but a thorough exploration is beyond the scope of this work.

The enormous variety of chemical compositions offered by the periodic table of elements has led us to further investigate several new stoichiometries with the CC setup, as described below. We wanted to know what the most extreme values of Δq are, what effects such a large Δq has on the crystal structure of OX and RED and what effects the charge transfer could have on the electronic and magnetic properties. The methodology employed here is identical as in (28,31,31,51) and its details are contained in SM.

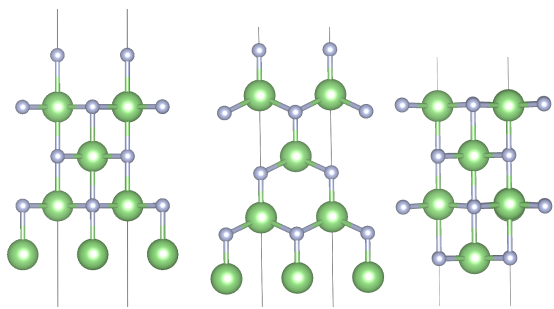

Figure 6. Structure of the Li | (LiF)$_3$ | F system: (left) pre-optimization, and (middle) following full optimization (projection on the *ac* plane). (right) Structure of the (LiF)$_4$ layer with the chemical formula equivalent to that of the Li | (LiF)$_3$ | F system. Li – large green, F – small gray balls.

**Alkali metal | alkali metal halide | halogen.** The first system calculated here is Li | (LiF)$_3$ | F. In other words, we bring the very reactive alkali metal lithium and the immensely reactive non-metal fluorine (as atomic radicals, (38)) into close proximity. The difference in chemical potentials is of course immense and such reagents would explode if they came into contact with each other. Here, however, we separate them by an inert LiF tri-layer that should not react with any of the RED/OX reagents. In this way, the explosion is kept "on a leash" in the computer (whether this sophisticated experiment can be done in the lab without getting your fingers ripped off is another question entirely). What is the end result when you bring these reagents together at a distance? Figure 6 illustrates the initial (before optimization of the geometry) and final (after optimization) structure. The associated numerical data for this and other systems are gathered in the Table 1.

Our starting structure features a fully optimized (LiF)$_3$ trilayer, with Li atoms placed on its one side, and F atoms on the other. The starting Li…F separation between Li atoms and the closest fluoride anions from the (LiF)$_3$ slab is 1.95 Å, while the separation between the F atoms and the nearest Li$^+$ cations from the slab is 2.07 Å. However, the optimized structure differs markedly from the starting one. First, the abovementioned distances drop to 1.75 Å and 1.79 Å, respectively. In this way, they become very similar to the closest Li…F separations within the (LiF)$_3$ slab of 1.75 Å, indicating that the nature of all these bonds is similar (and ionic). This implies that a substantial CT has taken place between Li and F atoms. Indeed, the Δq for this system is 0.31 e per Li or F atom as calculated using integration of electronic DOS (28).

The second difference between the starting geometry and the optimized one consists of immense puckering of the (LiF)$_3$ separator slab for the latter (Figure 6). This may be measured by the departure of the Li$^+$ cations from F$^-$ anions of 0.66 Å in the crystallographic *c* direction (the pre-optimization one being obviously null). This is of course associated with the appearance of the huge dipole moment within the separator layer, which attempts to counteract the dipole moment generated by the external Li$^{0.31+}$ and F$^{0.31-}$ species. This is another testimony to the large electric field which appears in the system due to the CT. Since the static dielectric constant of the LiF bulk crystal is quite large, around 9 (39), the thin (LiF)$_3$ slab is severely distorted in the electric field.

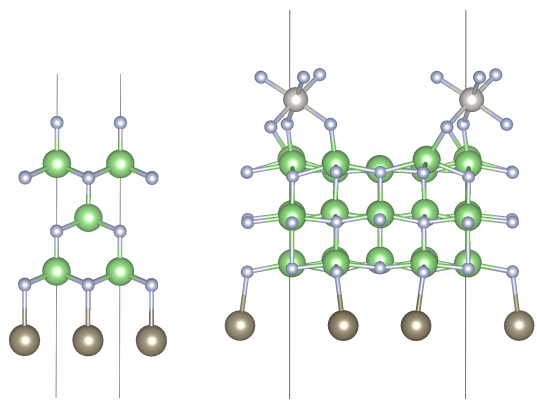

Figure 7. Optimized structure of the (left) Tl | (LiF)$_3$ | F and (Tl$_4$) | (Li$_8$F$_8$)$_3$ | F(PtF$_5$) (right). Li – large green, F – small gray, Pt – large gray, Tl – large brown balls.

As the optimized structure of Li | (LiF)$_3$ | F indicates the presence of the large electric field, it should exhibit propensity towards ion migration along the crystallographic *c* direction. While such migration is not possible in our computational experiment, due to integrity of the (LiF)$_3$ separator slab, yet one might anticipate that it could occur in the actual experiment due to imperfections and vacancies within that slab. Assuming that such migration could take place, it is interesting to ask what is the energetic gain connected with the formation of the (LiF)$_4$ four-layer slab with chemical formula equivalent to that of the Li | (LiF)$_3$ | F system (Figure 6). It turns out that

the energy gain is computed to be as large as 1.48 eV per LiF unit. This number testifies to how large is the driving force towards the ion diffusion, which may cancel the permanent dipole moment of the structure. This implies that CC structures which could be targeted in experiments should consist of extremely high-quality nanoscopic SEP layers to be metastable, at least at low temperatures.

**Metal | alkali metal halide | superhalogen.** The two other new systems worth looking at are Tl | (LiF)$_3$ | F and related (Tl$_4$) | (Li$_8$F$_8$)$_3$ | F(PtF$_5$) (Figure 7). In the latter, a single PtF$_6$ is present per four Tl atoms due to spatial constraints on the bulky molecular hexafluoride group in our periodic model.

By creating these systems, we have followed a classical approach of chemistry, where the presence of a strong Lewis acid (here: MF$_5$, M=Sb, Pt) enhances the oxidizing properties of an OX. Such tactics has been frequently utilized e.g. in inorganic fluorine chemistry (40,41,42). By attaching a Lewis acid to F atom, a so-called superhalogen is formed, i.e. a species with the electron affinity larger than that of the Cl atom (41). This in turn should increase the $E_V^{max}$–$E_C^{min}$ term and, consequently, also the Δq value.

Indeed, the calculated Δq values for these systems are 0.27 e and 0.88 e per one OX unit (F or PtF$_6$) for PtF$_5$–free and PtF$_5$–containing system, respectively. Clearly, the presence of a Lewis acid, PtF$_5$, enhances charge transfer per OX unit, in agreement with expectations. On the other hand, F atoms pack more efficiently on the LiF surface than the larger PtF$_6$ units. Obviously, both factors contribute to the chemical potential of the OX layer and influence the charge transferred.

Another interesting aspect of the CC may be seen for the (Tl$_4$) | (Li$_8$F$_8$)$_3$ | F(PtF$_5$) system (Figure 7), namely the symmetry breaking of the Tl sublattice. While for the optimized high-symmetry Tl | (LiF)$_3$ | F system the closest Tl…Tl separations are all equal to 3.95 Å, those for the (Tl$_4$) | (Li$_8$F$_8$)$_3$ | F(PtF$_5$) one they are 3.28 Å and 4.63 Å (both along the crystallographic ***a*** and ***b*** axes). The fact that these distances are, respectively, smaller and larger than those for the Tl | (LiF)$_3$ | F system suggests that some sort of Peierls distortion takes place. One could easily understand it within the Zintl-Klemm concept (2). Say, if Tl atoms were devoid of 1 e$^−$ each, they would constitute the (sp)$^2$ species; having two valence electrons, an atom may either form a molecule with a double element-element bond, an infinite polymer with two single bonds to each atom, or a cyclic species (likewise). In our case, although there is only partial (i.e. non-integer CT) yet Tl sublattice features distinct Tl$_4$ squares with four Tl…Tl bonds of 3.28 Å each, which are connected to other squares by weaker interactions (at 4.63 Å).

The symmetry breaking is supposed to be quite general feature for many CC systems. For example, the idealized high-symmetry Li | (LiF)$_3$ | F system discussed above is characterized by Δq close to 0.3 e per F atom. An educated guess dictates that if a sufficiently large and low-symmetry unit cell was constructed, this would supposedly lead to the lowest-energy solution with 3 F$^−$ anions and 7 F atoms rather than uniformly spread partial charge density on each F center (or some equivalent polyfluorides, e.g. 3 F$_3^−$ and 1 F, or 6 F$_3^−$ and 1 F$_2$). In such way, lots of electronic density may be removed from the Fermi level in agreement with the Maximum Hardness Principle (3). It is worth mentioning that a similar concept exists in condensed matter theory. High DOS at the Fermi level leads to a high electronic susceptibility (the opposite of hardness) which can promote a Stoner instability and drive the system to a state with a band gap (maximum hardness). Unfortunately, studying this follow-up "complex polyhalide" scenario *in silico* requires much more computational resources than those available to us at the moment.

**Metal | alkali metal halide | noble gas fluoride.** Since noble gas fluorides are known to be very strong oxidizers, we were tempted to study a system containing such OX species. We focused our attention on (Tl)$_3$ | (Li$_2$F$_2$)$_3$ | (KrF$_2$) (Figure 8), with three distinct Tl layers and a single KrF$_2$ molecule in the unit cell. Just like in the case of PtF$_6$, placing KrF$_2$ molecules at every rather than every second Li cation would cause spatial crowding and repulsion between them, so we follow a more realistic less-dense setting.

It turns out that the charge transferred in this system amounts to 0.28 e per one KrF$_2$ unit. This is quite substantial but similar to the value of 0.27 e previously discussed for the Tl | (LiF)$_3$ | F system. Just like for PtF$_6$, the weakness of using KrF$_2$ in CC setup is linked to its molecular nature which results in a rather small surface concentration. Thus, having studied the atomic (F) or molecular oxidizers (PtF$_6$, KrF$_2$) let us now turn to PbO$_2$, a known strong oxidizer, which forms an extended non-molecular solid (43).

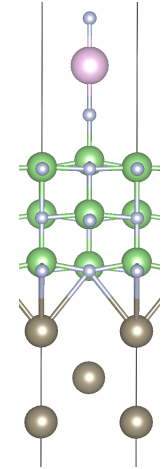

Figure 8. Optimized structure of Tl$_3$ | (Li$_2$F$_2$)$_3$ | (KrF$_2$). Li – large green, F – small gray, Kr – large pink, Tl – large brown balls.

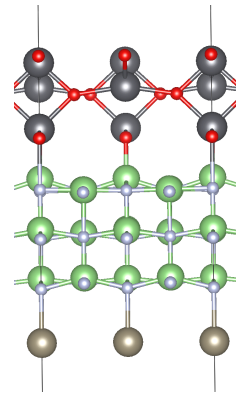

Figure 9. Optimized structure of (Tl$_2$) | (Li$_4$F$_4$)$_3$ | (PbO$_2$)$_3$. Li – large green, F – small gray, Pb – large black, O – small red, Tl – large brown balls.

**Metal | alkali metal halide | lead dioxide**. We have constructed a $(Tl_2) | (Li_4F_4)_3 | (PbO_2)_3$ system featuring a single layer of Tl, 3 layers of LiF separator, and three layers of $PbO_2$ (Figure 9). These components were selected according to their experimental lattice constants so that the sandwich system will not exhibit much strain within the *ab* plane and thus could in principle be achieved via epitaxy in experiment.

According to our calculations, this system exhibits Δq of 0.74 e per each $PbO_2$ unit. This is appreciable value, the largest among those discussed so far. We notice that – taking into consideration the stoichiometry of this system – it translates to the value of 1.11 h per each Tl atom. This is immense value since 1.00 hole in the electronic structure of Tl atom makes it similar to that of mercury! Clearly, the fact that strong OX, $PbO_2$, may densely feel the surface of a CC and in addition several layers of it may be placed together to increase the chemical potential, contribute to large CT achieved in this system. However, the drawback of $PbO_2$ as far as potential experiments are considered, is such that as a non-molecular OX, it cannot be deposited on LiF surface with a similar ease to a volatile molecular $KrF_2$.

The supplementary information (SI) contains additional data for the related $(Tl_4)_3 | (Li_8F_8)_3 | (Pb_2O_4)_5$ system, with the Δq of 0.57 e per each $PbO_2$ unit.

**Impact of the separator**. According to Eq.5, static dielectric constant of a separator should have substantial impact on charge transfer via $C_M$ factor, the larger the ε the larger the Δq. We have tested this idea by comparing OX=$MO_4$ (M=Ru, Os, Xe) and RED=Tl systems with two different separators, SEP = $Na_2MgF_4$ or $Sr_2TiO_4$ (Figure 10). Among these two isostructural double perovskites, $Na_2MgF_4$ does not have propensity to become FE, while $Sr_2TiO_4$ (which is essentially $SrTiO_3$ (44) with additional SrO layer) is a prototypical FE material.

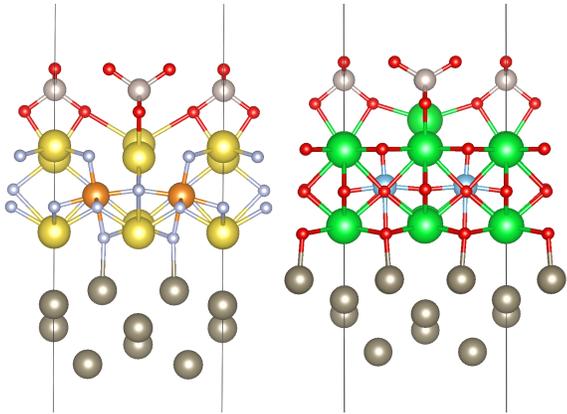

Figure 10. Optimized structure of (left) $(Tl_2)_3 | (Na_4Mg_2F_8) | (RuO_4)$ and (right) $(Tl_2)_3 | (Sr_4Ti_2O_8) | (RuO_4)$. Na – large yellow, F – small gray, Mg – orange, Ru – large gray, O – red, Sr – large green, Ti – light blue, Tl – large brown balls.

While at the first glance structure of both systems look similar to each other, yet the impact of the SEP manifests itself in a nearly 3-fold larger Δq of 1.74 e per Ru atom SEP = $Sr_2TiO_4$ than that of 0.61 e per Ru atom for SEP = $Na_2MgF_4$. Very similar ratios of Δq's were computed for these separators when $MO_4$ (M= Os, Xe) were used as OX (*cf.* SM and Table 1). Use of a FE material as a separator greatly enriches the possibilities to tune properties of a CC device via a judicious choice of its constituents, namely OX, RED and SEP.

Having gained some intuition as to how Δq may be manipulated via judicious selection of OX, RED and SEP constituents, let us now utilize this for doping of several important systems.

**Doping of the antiferromagnetic oxocuprate(II)**. SC occurs in a large number of oxocuprates including a h-doped $(Sr_{1-x}Ca_x)_{1-y}CuO_2$ (45). This so-called infinite-layer tetragonal system is of great importance since it does not have any heavy-transition metal charge reservoir layers; instead, it features the key constituent of any oxocuprate SC, namely the flat [$CuO_2$] sheet. If optimally doped, this system in the bulk may achieve the critical SC temperature of 110 K (46). Here, to avoid construction of huge supercells needed to describe $Sr_{1-x}Ca_x$ sublattice, we have selected its simpler $CaCuO_2$ analogue for the study. In Figure 11 we show the optimized structure of a monolayer of this compound in the antiferromagnetic state and placed on a LiF substrate together with its e- and h-doped versions. In Figure 12 we show the electronic density of states (DOS) of these three systems which reflects the contribution from Cu atoms only.

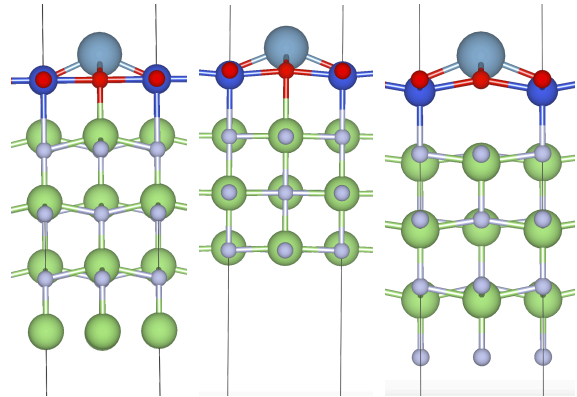

Figure 11. Optimized structure of (left) $(CaCuO_2) | (Li_2F_2)_3 | (Li_2)$, (center) $(CaCuO_2) | (Li_2F_2)_3 | \square$ where $\square$ stands for lack of any additional component, and (right) $(CaCuO_2) | (Li_2F_2)_3 | (F_2)$. Ca – large gray-blue, F – small gray, Li – large green, O – red, Cu – small blue balls. Smaller crystallographic cells (with 1 Cu atom) rather than magnetic unit cells (2 Cu atoms) are shown for clarity. Different directions of change transfer (e *vs.* h doping) is readily seen in the way the LiF layers of SEP are distorted.

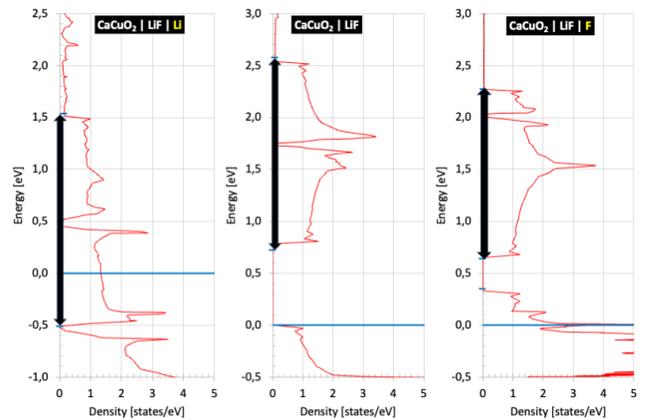

Figure 12. The calculated partial (atomic) electronic density of states coming from Cu for (left) $(CaCuO_2) | (Li_2F_2)_3 | (Li_2)$, (center) $(CaCuO_2) | (Li_2F_2)_3 | \square$ where $\square$ stands for lack of any additional component, and (right) $(CaCuO_2) | (Li_2F_2)_3 | (F_2)$. Horizontal blue line marks the position of the Fermi level for each system. Black arrows show the extent of the conduction band.

It is seen from Table 1 and Figure 12 that [CuO$_2$] sheet may be doped up to Δq=0.21 h and up to Δq=0.35 e when applying OX=F or RED=Li as redox-active constituents. This means significant depopulation of the valence band, or substantial population of the conduction (upper Hubbard) band, respectively. The computed doping limits exceed the value of ca. ±0.16 which is a typical optimum doping for most oxocuprates. Obviously, one might approach optimum values by either using of a thicker SEP layer (28), and/or by altering the chemical potential of the OX or RED layers. For example, using XeF$_2$ or KrF$_2$ as OX instead of F (while keeping the SEP thickness unchanged) one reaches much smaller Δq of 0.02 h. On the other hand, changing the stoichiometry of CaCuO$_2$ layer to isoelectronic Na$_2$CuO$_2$ while stoichiometry of the redox-active layers is preserved, results in smaller absolute values of Δq (Table 1). These examples demonstrate that one could manufacture a SC nano-device with the desired degree of charge transfer without the need of applying of a huge electric field in a field-effect transistor setup (25,26,27). Thus, the need of applying external voltage for sustaining of SC properties is eliminated; it is the chemical potential of a reactive near-by layer (OX or RED) which results in a permanent doping. Obviously, this is also true for the [AgF$_2$] layers which are isoelectronic to [CuO$_2$]$^{2-}$ ones (28). The doping is also manifested in the reduction of Cu(II) or Ag(II) magnetic moments (28).

**Doping of the cesium chloroaurate(I,III)**. Another interesting case is that of cesium chloroaurate, CsAuCl$_3$, a prototypical disproportionated (or charge density wave, CDW) system (47). Although the stoichiometry of this and related bromo- or iodoaurate systems might suggest the presence of Au(II) (isoelectronic to Cu(II) discussed above), yet in fact these systems feature an equimolar amount of Au(I) and Au(III) in their structure as manifested by very different Au-Cl bond lengths and coordination numbers at both Au sites. As such, they develop the gap at the Fermi level and are semiconducting. Here, we have studied a layer of CsAuCl$_3$ on Na$_2$MgF$_4$ substrate without and with additional OX or RED layers (Table 1). The corresponding partial electronic DOS for Au atoms is shown in the Figure 13.

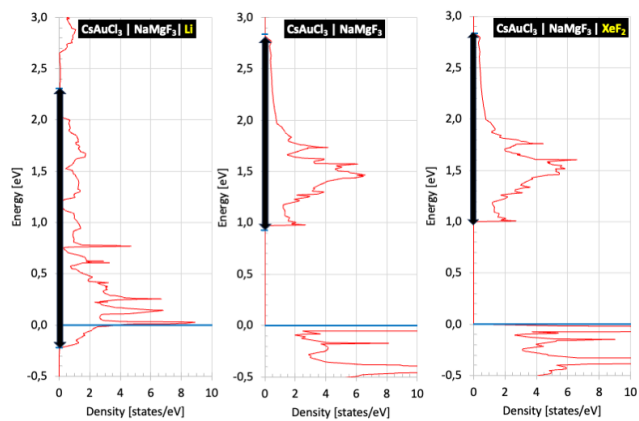

Figure 13. The calculated partial (atomic) electronic density of states coming from Au for (left) (Cs$_4$Au$_4$Cl$_{12}$) | (Na$_8$Mg$_4$F$_{16}$) | (Li$_4$), (center) (Cs$_4$Au$_4$Cl$_{12}$) | (Na$_8$Mg$_4$F$_{16}$) | □ where □ stands for lack of any additional component, and (right) (Cs$_4$Au$_4$Cl$_{12}$) | (Na$_8$Mg$_4$F$_{16}$) | (Xe$_4$F$_8$). Horizontal blue line marks the position of the Fermi level for each system.

Our calculations show that for RED=Li the resulting doping level to the conduction band is 0.10 e per Au. The electronic doping for RED=Li results in metallization of CsAuCl$_3$ and without the need of applying of an external pressure for this purpose (48). Interestingly, XeF$_2$ OX fails to withdraw any discernable electron density from the valence band (Δq = 0 h). This comes from the fact that in this case the key condition of functionality of a CC ($E_v^{max} < E_c^{min}$) is not fulfilled.

**Doping of other important families of systems**. Our additional results (not shown) indicate that similarly to oxocuprates(II), and to chloroaurates(I/III), one might also dope the oxobismuthates(III,V) (prototypical CDW precursors of SC) (49) and oxomanganites(III) (which are known to host giant magnetoresistance, GMR, upon doping) (17). Therefore, achieving SC/GMR in these parent systems could be possible without the need to modify the very layers hosting the desired property; instead, source of chemical potential in their vicinity could be employed.

**Selected parameters of CC systems** Table 1 gathers several key parameters of systems described in this work, namely: their stoichiometry, distance between OX and RED layers $d$, surface area of one unit cell of the CC, $A$, and a resulting charge transferred per 1 OX unit $\Delta q$.

It is readily seen from Table 1 that extreme value of Δq=1.74 e is obtained for (Tl$_2$)$_3$ | (Sr$_4$Ti$_2$O$_8$) | (RuO$_4$) system. Obviously, for each system studied the Δq may be easily decreased by either increasing the OX/RED separation, and/or by using a separator with a smaller dielectric constant. In any case, the buildup of electric charge implies the appearance of an appreciable electric field in these systems.

As already emphasized, the large values of the electric field present in the CC have implications for experiment, as they imply a possibly large ionic mobility. If, say, a fraction of Li$^+$ ions from the Li | (LiF)$_3$ | F system could diffuse to the F$^-$–rich external layer (which corresponds to the leak current of a physical capacitor), they would instantly reduce the charge transferred between the layers, and all other parameters of the CC with it. This suggests that only the highest quality, defect-free nano-structures manufactured by humans could exhibit really large values of Δq and all related properties. It is currently unknown what record values could be achieved in practice, but there are 3 orders of magnitude difference of the Δq between the only experimental manifestation of the CC so far (33) and the large theoretical values reported here.

**SC studies in a truncated CC setup**. As the inherent functionality of a CC consists of an appearance of metallic layers of partially doped OX or RED, it is natural to ask whether these layers could exbibit SC and what T$_C$ values they could achieve (28,31,32). This question is readily addressed for classical (BCS-type) superconductors, where T$_C$ can be rigorously computed from mathematical equations. However, the SC calculations are extremely CPU-demanding. Therefore, some studies use a truncated setup (Figure 5) to facilitate the computations (32). In such studies one uses an explicit partial e or h doping and doping level (δ) is a free parameter rather than the QM-calculated one.

We have used here a similar truncated setup here to study a potential SC in a doped single layer of NaCl, a Mg-Cu hydride (50), as well as of graphene (Figure 14). In the first case we have sandwiched the NaCl monolayer between two KMgF$_3$ ones, the KF layers being adjacent to the NaCl one. This allows for increase of the coordination numbers of Cl and Na ions to 6, just like in the bulk NaCl. In the case of Mg-Cu hydride and graphene, we did not use any adjacent supports at all, since such monolayers have a sufficient structural stiffness on their own.

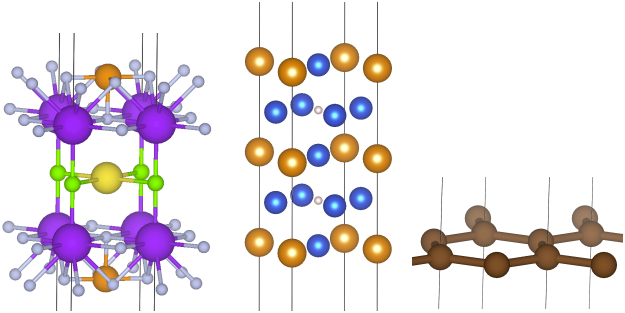

Figure 14. The crystal structures used for truncated CC studies of SC: (left) (KMgF$_3$) | (NaCl) | (KMgF$_3$), (middle) □ | (Mg$_3$Cu$_7$H$_4$) | □, and (right) □ | (C$_2$) | □, where □ stands for lack of any additional component. Na – large yellow, Cl – small green, K – large violet, Mg – orange, F – small gray, Cu – blue, H – small white, C – brown balls.

Interestingly, our calculations show that the NaCl monolayer may be doped up to 0.1 h per formula unit without undergoing any undesirable structural distortions. Such structure is predicted to be a SC with a $T_C$ of 68 mK. While this value may seem small, yet the possible appearance of SC is of interest since it involves introduction of holes to the Cl (2p) states ("metallic chlorine sublattice"). Instead, a 0.2 e doping leads to a SC with the computed $T_C$ value of 0.216 K (Figure 15).

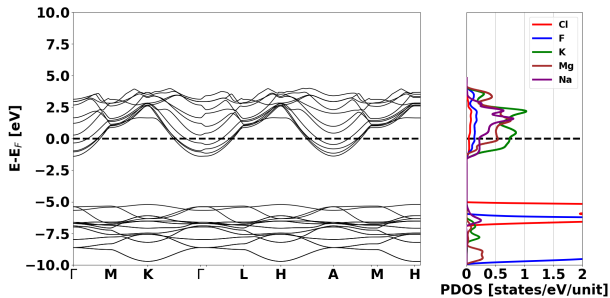

Figure 15. The electronic band structure of (KMgF$_3$) | (NaCl) | (KMgF$_3$) setup at 0.2 e doping.

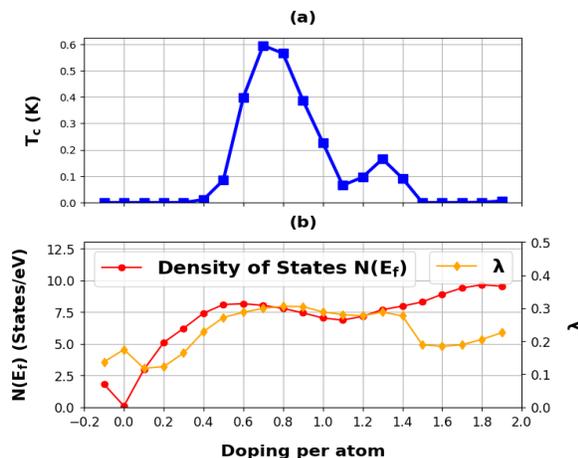

Figure 16. (a) $T_C$ of graphene as a function of doping per C atom. Negative and positive doping values correspond to hole and electron doping, respectively. (b) Corresponding density of states N(E$_f$) at the Fermi level and electron-phonon coupling constant λ as functions of doping. Reproduced with permission from (51).

The case of graphene is also very interesting. According to our calculations, range of doping to graphene is very asymmetrical with respect to type of the charge carriers: graphene withstands up to 0.1 h doping but as much as 1.9 e doping per C atom (sic!) without structural distortions (Figure 16) (51). It turns out that the calculated $T_C$ value varies from up to ca. 0.6 K for 0.7 e doping level, with the second distinct maximum of ca. 0.2 K for 1.3 e doping one (Figure 17). Again, these values may seem small, but so far, the highest $T_C$ value achieved for two graphene layers at a "magic angle" of 1.1° with respect to one another was 1.7 K (52).

Last but not the least, the Mg-Cu hydride sandwich studied here, is also predicted to exhibit SC. This two-dimensional (2D) system is composed of alternating MgCu and Cu$_2$H units, just like its parent MgCu$_3$H perovskite in the bulk (50). Our results show that such sandwich sustains up to 10% e- and up to 36% h-doping without structural collapse. However, the $T_C$ dependence in the function of the doping level is quite complex (Figure 17).

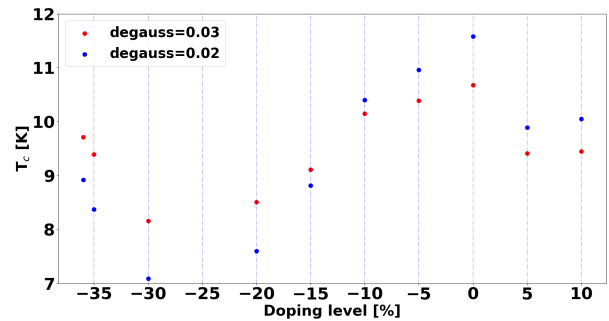

Figure 17. The $T_C$ dependence on the doping level for the □ | (Mg$_3$Cu$_7$H$_4$) | □ sandwich calculated with two distinct degauss values of 0.02 or 0.03 (cf. SM for details). Negative and positive doping values correspond to hole and electron doping, respectively.

The calculated $T_C$ value ranges from ca. 7 to ca. 12 K (for the degauss value of 0.02) for the entire dynamic stability domain, with the maximum value of 11.6 K at the null doping level (cf. SM for more SC parameters). Although the $T_C$ drops for both small h- and e-doping level around the maximum, yet it rises again at larger doping levels when dynamic instability regime is approached. This indeed could be expected for the BCS-type SC driven by a softening phonon modes (32).

**Summary and prospect.** This work presents a detailed computational investigation into the doping of two-dimensional materials, with a chemical capacitor (CC) architecture. We showed that by strategically integrating redox-active layers (OX/RED) with appropriate separator layers, substantial and tunable electronic doping can be achieved without applying an external electric field, thus enabling the permanent induction of metallic, superconducting (SC), or other symmetry broken states. The degree of doping of a RED layer (quantified as *Δq*) may be finely tuned by judiciously choosing the OX layer (or *vice versa*), the SEP layer and its thickness.

Particularly interesting realization of CC in experiment would be to tune the doping to regions of high density of states where other broken symmetry states may arise as ferromagnetism. Another interesting prospect is to study the effect of spin orbit coupling. A strong electric field is known to produce a Rashba splitting of bands which enables new functionalities as, for example, spin to charge conversion (53).

Despite these promising prospects, the realization of such functionalities in practice will require manufacturing nanostructures of very good quality to limit ionic mobility and minimize defect-driven leakage. However, the experimental realization of similar architectures as previous CC (33), the two-dimensional electron gas in polar interfaces (34,35,36) and Janus materials (37), testify the degree of sophistication that current nanotechnologies had achieved and pave the way to the experimental realization of the present CC architectures.

## ASSOCIATED CONTENT

**Supporting Information**. contents This material is available free of charge via the Internet at http://pubs.acs.org. The SI contains cif files of optimized crystal structures, describes computational methodology, method of estimation of charge transfer, lists SC parameters for two systems, and cites additional references from (54) to (71).

## AUTHOR INFORMATION

**Corresponding Authors**

* l.wolanski@cent.uw.edu.pl, w.grochala@cent.uw.edu.pl

**Author Contributions**

The manuscript was written through contributions of all authors. D.C. and P.S. had equal contribution.

**Funding Sources**

W.G. is grateful to Polish National Science Center for project OPUS Chem-Cap (2021/41/B/ST5/00195).

**Notes**

This paper is dedicated *in memoriam* Prof. Bogumił Jeziorski (1947–2023).

## ACKNOWLEDGMENT

Computations were performed at the Interdisciplinary Center for Mathematical and Computational Modeling of Warsaw University (grant SAPPHIRE GA83-34) and at the Wrocław Centre for Networking and Supercomputing (grant No. 484).

## ABBREVIATIONS

CC, chemical capacitor; ECC, electrochemical cell; PC, physical capacitor; MV, mixed valence; IV, intermediate valence; NPJ, n-p junction; SEP, separator; OX, oxidizer; RED, reductant; $E_F$, energy of the Fermi level; $E_V^{max}$, energy of the top of the valence band; $E_C^{min}$, energy of the bottom of the conduction band; μ, chemical potential; V, electric potential; ε, dielectric constant (relative electric permittivity); $\varepsilon_0$, electric permittivity of vacuum; δ, doping level; DNA, deoxyribonucleic acid; RNA, ribonucleic acid; FET, field effect transistor; CPU, central processing unit; CCD, charge coupled device; DFT, density functional theory; CT, charge transfer; q, amount of electric charge; SC, superconductor; GMR, giant magnetoresistance; CDW, charge density wave; 2D, two-dimensional.

## REFERENCES

(1) https://en.wikipedia.org/wiki/Law_of_definite_proportions; accessed Jan 25, 2025.
(2) E. Zintl, J. Goubeau, W. Dullenkopf, *Z. Phys. Chem. A* 1931, 1–46.
(3) W. Grochala, *Phys. Chem. Chem. Phys.* 2017, **19**, 30964–30983; W. Grochala, *Phys. Chem. Chem. Phys.* 2017, **19**, 30984–31006.
(4) W. Hume-Rothery, Electrons, atoms, metals and alloys. Dover, New York, USA, 1963.
(5) D. Jezierski, K. Koteras, M. Domański, P. Połczyński, Z. Mazej, J. Lorenzana, W. Grochala, *Chem. Eur. J* 2023, **29**, e202301092.
(6) A. Bhaumik, R. Sachan, J. Narayan, *ACS Nano* 2017, **11**, 5351–5357.
(7) M. A. Bredig, J. W. Johnson, W. T. Smith, *J. Amer. Chem. Soc.* 1955, **77**, 307–312.
(8) S. J. Yosim, A. J. Darnell, W. G. Gehman, S. W. Mayer, *J. Phys. Chem.* 1959, **63**, 230–233.
(9) T. Graham, *Proc. Royal. Soc.* 1868, 212–220.
(10) J. B. Goodenough. *Phys. Rev. B*, 1972, **5**, 2764–2774.
(11) F. Rivadulla, J. Fernández-Rossier, M. García-Hernández, M. A. López-Quintela, J. Rivas, J. B. Goodenough, *Phys. Rev. B* 2007, **76**, 1–6.
(12) K. Fredenhagen, G. Cadenbach, *Z. Anorg. Allg. Chem.* 1926, **158**, 249–265.
(13) M. M. Thackeray, W. I. F. David, P. G. Bruce, J. B. Goodenough, *Mater. Res. Bull.* 1983, **18**, 461–472.
(14) J. G. Bednorz, K. A. Müller, *Z. Phys. B*, 1986, **64**, 189–193.
(15) S. Yamanaka, K. -I. Hotehama, H. Kawaji, *Nature* 1998, **392**, 580–582.
(16) First prepared by F. Wöhler in 1823, as described in: P. Hagenmuller, "Chapter 50: Tungsten bronzes, vanadium bronzes, and related compounds", Comprehensive Inorganic Chemistry, Vol. 4, Pergamon, pp. 541–605, 1973.
(17) S. Jin, T. H. Tiefel, M. McCormack, R. A. Fastnacht, R. Ramesh, L. H. Chen, *Science* 1994, **264**, 413–415.
(18) A. K. Cheetham, B. E. F. Fender, M. J. Cooper, *J. Phys. C: Solid State Phys.*, 1971, **4**, 3107–3121, and references within.
(19) C. K. Chiang, M. A. Druy, S. C. Gau, A. J. Heeger, E. J. Louis, A. G. MacDiarmid, Y. W. Park, H. Shirakawa, *J. Amer. Chem. Soc.* 1978, **100**, 1013–1015.
(20) M. Somayazulu, M. Ahart, A. K. Mishra, Z. M. Geballe, M. Baldini, Y. Meng, V. V. Struzhkin, R. J. Hemley, *Phys. Rev. Lett.* 2019, **122**, 027001.
(21) A. P. Drozdov, P. P. Kong, V. S. Minkov, S. P. Besedin, M. A. Kuzovnikov, S. Mozaffari, L. Balicas, F. F. Balakirev, D. E. Graf, V. B. Prakapenka, E. Greenberg, D. A. Knyazev, M. Tkacz, M. I. Eremets, *Nature*, 2019, **569**, 528–531.
(22) J. -W. Yeh, S. -K. Chen, S. -J. Lin, J. -Y. Gan, T. -S. Chin, T. -T. Shun, C. -H. Tsau, S. -Y. Chang, *Adv. Eng. Mater.* 2004, **6**, 299–303, and references within.
(23) C. M. Rost, E. Sachet, T. Borman, A. Moballegh, E. C. Dickey, D. Huo, J. L. Jones, S. Curtarolo, J. -P. Maria, *Nat. Commun.*, 2015, **6**, 848.
(24) P. Piekarz, K. Parlinski, A. M. Oleś, *Phys. Rev. Lett.* 2006, **97** 156402.
(25) G. Dubuis, A. T. Bollinger, D. Pavuna, I. Bozovic, *J. Appl. Phys.* 2012, **111**, 112632.
(26) J. -F. Ge, Z. -L. Liu, C. Liu, C. -L. Gao, D. Qian, Q. -K. Xue, Y. Liu, J. -F. Jia, *Nat. Mater.* 2015, **14**, 285–289.
(27) P. Liu, B. Lei, X. H. Chen, L. Wang, X. L. Wang, *Nat. Rev. Phys.* 2022, **4**, 336–352.
(28) A. Grzelak, J. Lorenzana, W. Grochala, *Angew. Chem. Int. Ed. Engl.* 2021, **60**, 13892–13895.
(29) W. Grochala, *Wiadom. Chem.* 2021, **75**, 755–769 (in Polish).
(30) W. Grochala, R. Hoffmann, *Angew. Chem. Int. Ed. Engl.* 2001, **40**, 2742–2781.
(31) D. Jezierski, A. Grzelak, L. Xiao-Qiang, S. K. Pandey, M. N. Gastiasoro, J. Lorenzana, J. Feng, W. Grochala, *Phys. Chem. Chem. Phys.* 2022, **24**, 15705–15717.
(32) P. Szkudlarek, C. Renskers, R. Margine, W. Grochala, *ChemPhysChem* 2025, **26**, e202500013.

(33) R. Larciprete, P. Lacovig, F. Orlando, M. Dalmiglio, L. Omiciuolo, A. Baraldi, S. Lizzit, *Nanoscale* 2015, **7**, 12650–12658.
(34) A. Ohtomo, H. Y. Hwang, *Nature*, 2004, **427**, 423–427.
(35) S. Thiel, G. Hammerl, A. Schmehl, C. W. Schneider, J. Mannhart, *Science*, 2006, **313**, 1942–1945.
(36) N. Nakagawa, H. Y. Hwang, D. A. Muller, *Nat. Mater.* 2006, **5**, 204–209.
(37) L. Zhang, Z. Yang, T. Gong, R. Pan, H. Wang, Z. Guo, H. Zhang, X. Fu, *J. Mater. Chem. A*, 2020, **8**, 8813–8830.
(38) N. Bartlett, *Proc. Chem. Soc. London* 1962, 218–218; J. L. Weeks, C. L. Chernick, M. S. Mateson, *J. Amer. Chem. Soc.* 1962, **84**, 4612–4613.
(39) C. Andeen, J. Fontanella, D. Schuele, *Phys. Rev. B*, 1970, **2**, 5068–5073.
(40) M. Lerner, R. Hagiwara, N. Bartlett, *J. Fluor. Chem.* 1992, **57**, 1–13.
(41) X. B. Wang, C. F. Ding, L. -S. Wang, A. I. Boldyrev, J. Simons, *J. Chem. Phys.* 1999, **110**, 4763–4771.
(42) R. Craciun, D. Picone, T. R. Long, S. Li, D. A. Dixon, K. A. Peterson, K. O. Christe, *Inorg. Chem.* 2010, **49**, 1056–1070.
(43) D. O. Scanlon, A. B. Kehoe, G. W. Watson, M. O. Jones, W. I. F. David, D. J. Payne, R. G. Egdell, P. P. Edwards, A. Walsh, *Phys. Rev. Lett.* 2011, **107**, 246402.
(44) R. Migoni, H. Bilz, D. Bauerle, *Phys. Rev. Lett.* 1976, **37**, 1155–1158.
(45) T. Siegrist, S. M. Zahurak, D. W. Murphy, R. S. Roth, *Nature* 1988, **334**, 231–232.
(46) Z. Hiroi, M. Azuma, M. Takano, Y. Takeda, *Physica C: Supercond.* 1993, **208**, 286–296.
(47) N. Elliott, L. Pauling, *J. Amer. Chem. Soc.* 1938, **60**, 1846–1851.
(48) S. S. Hafner, N. Kojima, J. Stanek, L. Zhang, *Phys. Lett. A*, 1994, **192**, 385–388; S. B. Wang, A. F. Kemper, M. Baldini, M. C. Shapiro, S. C. Riggs, Z. Zhao, Z. Liu, T. P. Devereaux, T. H. Geballe, I. R. Fisher, W. L. Mao, *Phys. Rev. B* 2014, **89**, 245109.
(49) A. W. Sleight, J. L. Gillson, P. E. Bierstedt, *Solid State Commun.* 1975, **17**, 27–28.
(50) C. Tian, Y. He, Y. -H. Zhu, J. Du, S. M. Liu, W. -H. Guo, H. -X. Zhong, J. Lu, X. Wang, J. J. Shi, *Adv. Funct. Mater.*, 2024, **34**, 2304919.
(51) D. Ciszewski, W. Grochala, *Chem. Commun.*, in press 2025, https://doi.org/10.1039/D5CC02642C.
(52) Y. Cao, V. Fatemi, S. Fang, K. Watanabe, T. Taniguchi, E. Kaxiras, P. Jarillo-Herrero, *Nature* 2018, **556**, 43–50.
(53) J. C. Rojas Sánchez, L. Vila, G. Desfonds, S. Gambarelli, J. P. Attané, J. M. De Teresa, C. Magén, A. Fert, *Nat. Commun.* 2013, **4**, 2944.
(54) J. P. Perdew, K. Burke, M. Ernzerhof, *Phys. Rev. Lett.*, 1996, **77**, 3865–3868.
(55) J. P. Perdew, K. Burke, M. Ernzerhof, *Phys. Rev. Lett.*, 1997, **78**, 1396.
(56) G. I. Csonka, J. P. Perdew, A. Ruzsinszky, P. H. T. Philipsen, S. Lebègue, J. Paier, O. A. Vydrov, J. G. Ángyán, *Phys. Rev. B - Condens. Matter Mater. Phys.*, 2009, **79**, 155107.
(57) P. E. Blöchl, *Phys. Rev. B*, 1994, **50**, 17953–17979.
(58) G. Kresse, D. Joubert, *Phys. Rev. B*, 1999, **59**, 1758–1775.
(59) Vienna Ab initio Simulation Package, https://www.vasp.at.
(60) S. Grimme, J. Antony, S. Ehrlich, H. Krieg, *J. Chem. Phys.*, 2010, **132**, 154104.
(61) V. I. Anisimov, J. Zaanen, O. K. Andersen, *Phys. Rev. B*, 1991, **44**, 943–954.
(62) E. R. Ylvisaker, W. E. Pickett, K. Koepernik, *Phys. Rev. B - Condens. Matter Mater. Phys.*, 2009, **79**, 35103.
(63) B. Himmetoglu, A. Floris, S. De Gironcoli, M. Cococcioni, *Int. J. Quantum Chem.*, 2014, **114**, 14–49.
(64) P. Giannozzi, S. Baroni, N. Bonini, M. Calandra, R. Car, C. Cavazzoni, D. Ceresoli, G. L. Chiarotti, M. Cococcioni, I. Dabo, A. Dal Corso, S. De Gironcoli, S. Fabris, G. Fratesi, R. Gebauer, U. Gerstmann, C. Gougoussis, A. Kokalj, M. Lazzeri, L. Martin-Samos, N. Marzari, F. Mauri, R. Mazzarello, S. Paolini, A. Pasquarello, L. Paulatto, C. Sbraccia, S. Scandolo, G. Sclauzero, A. P. Seitsonen, A. Smogunov, P. Umari, R. M. Wentzcovitch, *J. Phys. Condens. Matter*, 2009, **21**, 395502.
(65) P. Giannozzi, O. Andreussi, T. Brumme, O. Bunau, M. Buongiorno Nardelli, M. Calandra, R. Car, C. Cavazzoni, D. Ceresoli, M. Cococcioni, N. Colonna, I. Carnimeo, A. Dal Corso, S. De Gironcoli, P. Delugas, R. A. Distasio, A. Ferretti, A. Floris, G. Fratesi, G. Fugallo, R. Gebauer, U. Gerstmann, F. Giustino, T. Gorni, J. Jia, M. Kawamura, H. Y. Ko, A. Kokalj, E. Kücükbenli, M. Lazzeri, M. Marsili, N. Marzari, F. Mauri, N. L. Nguyen, H. V. Nguyen, A. Otero-De-La-Roza, L. Paulatto, S. Poncé, D. Rocca, R. Sabatini, B. Santra, M. Schlipf, A. P. Seitsonen, A. Smogunov, I. Timrov, T. Thonhauser, P. Umari, N. Vast, X. Wu, S. Baroni, *J. Phys. Condens. Matter*, 2017, **29**, 465901.
(66) D. R. Hamann, *Phys. Rev. B - Condens. Matter Mater. Phys.*, 2013, **88**, 85117.
(67) M. Methfessel, A. T. Paxton, *Phys. Rev. B*, 1989, **40**, 3616–3621.
(68) S. Baroni, S. De Gironcoli, A. Dal Corso, P. Giannozzi, *Rev. Mod. Phys.*, 2001, **73**, 515–562.
(69) P. B. Allen, R. C. Dynes, *Phys. Rev. B*, 1975, **12**, 905–922.
(70) K. Momma, F. Izumi, *J. Appl. Crystallogr.*, 2008, **41**, 653–658.
(71) H. T. Stokes, D. M. Hatch, *J. Appl. Crystallogr.*, 2005, **38**, 237–238.

**Table 1. Stoichiometry of the RED | SEP | OX chemical capacitors, distance between OX and RED layers (d), surface area of one unit cell of the CC (A), charge transferred per 1 OX unit (Δq). NA – not applicable.**

| Stoichiometry | d[$] [Å] | A [Å$^2$] | Δq [e] |
|---|---|---|---|
| Li | (LiF)$_3$ | F | 9.036 | 7.81 | 0.307 |
| Tl | (LiF)$_3$ | F | 9.620 | 7.81 | 0.270 |
| (Tl$_4$) | (Li$_8$F$_8$)$_3$ | F(PtF$_5$) | 8.92 | 62.52 | 0.877 |
| (Tl)$_3$ | (Li$_2$F$_2$)$_3$ | (KrF$_2$) | 9.104 | 15.63 | 0.282 |
| (Tl$_2$) | (Li$_4$F$_4$)$_3$ | (PbO$_2$)$_3$ | 9.409 | 62.52 | 0.744 |
| (Tl$_4$)$_3$ | (Li$_8$F$_8$)$_3$ | (Pb$_2$O$_4$)$_5$ | 9.274 | 62.52 | 0.566 |
| (Tl$_2$)$_3$ | (Sr$_4$Ti$_2$O$_8$) | (KrF$_2$) | 8.736 | 15.07 | 0.611 |
| (Tl$_2$)$_3$ | (Sr$_4$Ti$_2$O$_8$) | (OsO$_4$) | 8.447 | 60.28 | 1.268 |
| (Tl$_2$)$_3$ | (Sr$_4$Ti$_2$O$_8$) | (RuO$_4$) | 8.462 | 60.28 | 1.738 |
| (Tl$_2$)$_3$ | (Sr$_4$Ti$_2$O$_8$) | (XeO$_4$) | 8.620 | 60.28 | 1.650 |
| (Tl$_2$)$_3$ | (Na$_4$Mg$_2$F$_8$) | (OsO$_4$) | 8.430 | 60.77 | 0.464 |
| (Tl$_2$)$_3$ | (Na$_4$Mg$_2$F$_8$) | (RuO$_4$) | 8.290 | 60.77 | 0.608 |
| (Tl$_2$)$_3$ | (Na$_4$Mg$_2$F$_8$) | (XeO$_4$) | 8.610 | 60.77 | 0.721 |
| ☐ | (Na$_8$Mg$_4$F$_{16}$) | (Cs$_4$Au$_4$Cl$_{12}$) | NA | 60.77 | 0 |
| (Li$_4$) | (Na$_8$Mg$_4$F$_{16}$) | (Cs$_4$Au$_4$Cl$_{12}$) | 8.085 | 60.77 | 0.097 |
| (Li$_4$)$_3$ | (Na$_8$Mg$_4$F$_{16}$) | (Cs$_4$Au$_4$Cl$_{12}$) | 7.993 | 60.77 | 0.092 |
| (Cs$_4$Au$_4$Cl$_{12}$) | (Na$_8$Mg$_4$F$_{16}$) | (Xe$_4$F$_8$) | 8.104 | 60.77 | 0 |
| ☐ | (Li$_2$F$_2$)$_3$ | (CaCuO$_2$)* | NA | 31.26 | 0 |
| (Li$_2$) | (Li$_2$F$_2$)$_3$ | (CaCuO$_2$)* | 8.800 | 31.26 | 0.349 |
| (CaCuO$_2$) | (Li$_2$F$_2$)$_3$ | (XeF$_2$)* | 8.360 | 31.26 | 0.016 |
| (CaCuO$_2$) | (Li$_2$F$_2$)$_3$ | (KrF$_2$)* | 8.410 | 31.26 | 0.022 |
| (CaCuO$_2$) | (Li$_2$F$_2$)$_3$ | (F$_2$)* | 8.590 | 31.26 | 0.104[&] |
| ☐ | (Li$_2$F$_2$)$_3$ | (Na$_2$CuO$_2$)^ | NA | 15.63 | 0 |
| (Li$_2$) | (Li$_2$F$_2$)$_3$ | (Na$_2$CuO$_2$)^ | 8.548 | 15.63 | 0.045 |
| (Na$_2$CuO$_2$) | (Li$_2$F$_2$)$_3$ | (XeF$_2$)^ | 8.836 | 15.63 | 0.058 |
| (Na$_2$CuO$_2$) | (Li$_2$F$_2$)$_3$ | (KrF$_2$)^ | 8.835 | 15.63 | 0.011 |
| (Na$_2$CuO$_2$) | (Li$_2$F$_2$)$_3$ | (F$_2$)^ | 8.997 | 15.63 | 0.031[&] |

[$] The distance between the layers is defined as the closest separation between the redox active-components of the OX and RED layers

* These systems are antiferromagnetic hence they required a sqrt2 x srt2 x 1 supercell for calculations; data shown here were reduced to a twice smaller crystallographic unit cell

^ These systems are calculated to be ferromagnetic hence the magnetic cell corresponds to the crystallographic unit cell

[&] For these systems doping level to Cu (discussed in text) is twice larger than doping level to OX=F due to 1:2 stoichiometry of Cu:F

**Graphical Abstract.** A detailed computational study into the doping of two-dimensional materials, with a chemical capacitor architecture, is presented. Strategic integration of redox-active layers with appropriate separator layers, allows for achieving of substantial and tunable electronic doping without applying an external electric field, thus enabling the permanent induction of metallic or super-conducting states. The degree of doping may be finely tuned by judiciously choosing the chemical stoichiometry of the layers and separator's thickness.

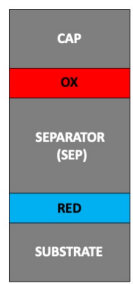
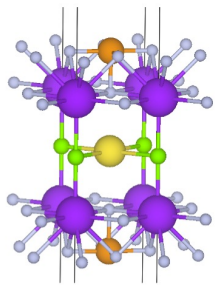

CHEMICAL CAPACITOR:

- 2D SYSTEMS
- TUNABLE CHARGE TRANSFER
- METALLICITY
- SUPERCONDUCTIVITY

# Supplementary Material

# Chemical capacitor – its concept, functionalities and limits

Łukasz Wolański[1,*], Dawid Ciszewski[1], Piotr Szkudlarek[1], José Lorenzana, and Wojciech Grochala[1,*]

[1]Centre of New Technologies, University of Warsaw, Żwirki i Wigury 93, 02089, Warsaw, Poland
[2]ISC-CNR and Department of Physics, Sapienza University of Rome, Piazzale Aldo Moro 2, 00185, Rome, Italy

## Table of Contents

# Supplementary Material

## 1. Computational details

Computational exploration of investigated structures was done by two-step approach. Firstly, equilibrium structures were obtained. In the second step, they were used in a static (single-point) calculations of electronic density of states (DOS). All these computations were conducted at DFT level of theory with Perdew-Burke-Ernzerhof (PBE) exchange-correlation functional[1,2] revised for solids (PBEsol)[3]. The projector-augmented-wave method[4] with appropriate pseudopotentials[5] from v.54 dataset were used as implemented in utilized by us VASP 5.4.4 code[6]. The cut-off energy of the plane wave basis set was equal to 850 eV with a self-consistent-field convergence criterion of $1·10^{-6}$ eV. All calculation were performed for density of the k-point grid of 0.07 Å$^{-1}$ and with accurate precision. Van der Waals dispersion correction was approximately taken into consideration by DFT-D3 approach of Grimme[7]. In the case of copper compounds, their magnetic nature was taken into account by DFT+U method[8–10] (U = 5.5 eV, J = 1 eV). We have found out that rigorous application of the periodic boundary conditions for the electric field in the *c* crystallographic direction does not lead to very different results for the transferred charge as compared to the scenario when this condition was omitted.

At this point we would like to point out, that geometry optimization computational approach appropriate for chem-cap structures had to take into account its two-dimensionality. In the case of calculations conducted with the three-dimensional-periodic methodology, subsequent chem-cap planes had to be separated to each other with a suitably large vacuum slab. In practice, this meant that the first step was to fully optimize (ISIF=3) the equilibrium structures of used separators as a 3D solids. Then, for each chem-cap the oxidant and reductant were appropriately applied to the separator layer. Such obtained cell had two dimensions, which lengths come from calculations for the 3D separator, and the third dimension (height) which was set as large enough to ensure planes separation. Geometry optimization of all such systems were performed with fixed shape and volume of cell (ISIF=2). Such scheme was applied for most structures studied.

Calculations of the electronic DOS, phononic DOS, and critical temperatures $T_c$ in the truncated CC setup were carried out using the Quantum ESPRESSO (QE) suite[11,12]. Perdew-Burke-Ernzerhof (PBE) optimized norm-conserving Vanderbilt (ONCV) pseudopotentials from the Pseudo Dojo library[13] were used as a basis for the density functional theory calculations. A plane-wave kinetic energy cutoff of 100 Ry for the wavefunctions and 400 Ry for the charge density and potential was used for all systems. Brillouin-zone integration involved a 18x18x1 Γ-centered k-mesh with a Methfessel-Paxton smearing width[14] of 0.02 Ry. Since our samples are 2D materials, the out-of-plane direction was sampled with only one k-point, and a vacuum slab of at least 10 Å was included. Structural optimization was considered converged when the total energy and atomic forces per atom reached tolerances of $10^{-6}$ Ry and $10^{-4}$ Ry/Å, respectively. The vacuum spacing was kept constant. Both graphene and NaCl systems were electron and hole doped until dynamic instability (i.e., the appearance of imaginary phonon modes) was reached. The dynamical matrices and linear variation of the self-consistent potential were computed using density-function perturbation theory (DFPT)[15] on the irreducible set of a regular 6x6x1 mesh. For each q-point, the electron-phonon matrix elements were calculated on a denser 54x54x1 k-mesh. The Eliashberg spectral function was calculated for a set of broadening ranging from 0 to 0.05 Ry. The superconducting critical temperature was determined using the Allen-Dynes modified McMillan formula[16], assuming a Coulomb pseudopotential of µ = 0.10 and at degauss value of 0.03. The Eliashberg spectral function ($α^2F$), electron-phonon coupling strength (λ), logarithmic average phonon frequency ($ω_{log}$), and superconducting temperature ($T_c$) formulas are shown below.

$$\alpha^2 F(\omega) = \frac{1}{N_F} \int \frac{\mathrm{d}k\mathrm{d}q}{\Omega_{BZ}^2} \sum_{mn\nu} |g_{mn\nu}(k,q)|^2 \delta(\epsilon_{nk} - \epsilon_F) \delta(\epsilon_{mk+q} - \epsilon_F) \delta(\hbar\omega - \hbar\omega_{q\nu})$$

$$\lambda = 2 \int_0^\infty \frac{\alpha^2 F(\omega)}{\omega} d\omega$$

$$\omega_{log} = \exp\left[\frac{2}{\lambda} \int_0^\infty d\omega \frac{\alpha^2 F(\omega)}{\omega} \log \omega\right]$$

$$\omega_2^2 = \frac{2}{\lambda} \int \omega \alpha^2 F(\omega) d\omega$$

The Eliashberg spectral function ($\alpha^2F(\omega)$), the electron phonon coupling strength ($\lambda$), the logarithmic frequency ($\omega_{\log}$), and $\omega_2$ are used to calculate the critical temperature through the modified Allen-Dynes modified McMillan formula:

$$k_B T_c = \frac{\hbar \omega_{\log}}{1.2} \exp\left[-\frac{1.04(1+\lambda)}{\lambda - \mu^*(1+0.62\lambda)}\right]$$

The $T_C$ calculations for the $Mg_3Cu_7H_4$ hydride were performed using identical procedure and the same underlying physics equations but instead of employing the lambda.x code of QE they used a matdyn.x one.

All structure visualizations in this study were obtained with the VESTA software[17]. The symmetry of obtained equilibrium structures was determined with FINDSYM 7.1.3[18].

## 2. Charge Transfer Calculation Methodology

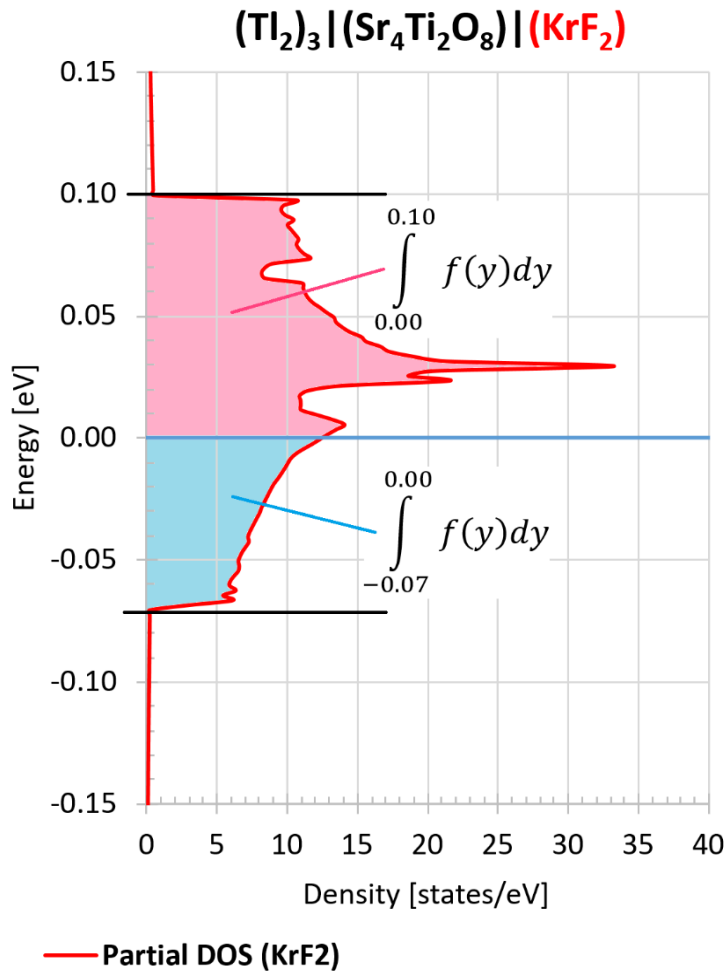

One of the key values determined for investigated systems is the value of charge transfer that occur in each of the studied chemical capacitors (chem-cap). Our approach to obtain its value base on partial DOS computed for the oxidizer layer (or simply for the metal center of this layer). In principle we avoid running into diverse electronic population schemes but rather we determine the charge transferred during a **redox process**, which is much simpler to calculate.

The oxidizer is as an electron acceptor. Typical result of its interaction with the reductant is filling electron shells/bands, that were empty as long as this layer was isolated from the other chem-cap parts. It is therefore crucial to examine such a partially occupied electron shell through which the Fermi level passes or are the nearest to it.

In practice, it is necessary to determine the energy limits in which this acceptor (conduction) band lies (which is only sometimes tricky) and then calculate the area of this band lying below and above the Fermi level (see Fig. above). In both cases integrals are calculated with trapezoidal rule approximation.

In the analyzed example (see Fig. above), the occupied part of the band (below the Fermi level) has an area of 0.571, while the unoccupied part has an area of 1.298. In total, this gives 1.869. This band accommodates in fact exact two electrons. The deviations of obtained data are results of inaccuracy of the DOS calculation method itself. Since they mainly concern the absolute values themselves, and not their mutual relationship, the charge that found itself in this orbital can be calculated proportionally: 0.571*(2.0/1.869). This gives a final charge of 0.611 e. Since this orbital would be empty for isolated oxidizer layer, this value corresponds to the transferred charge per 1 Kr atom.

## 3. CIF files of optimized structures.

**#LiF(2D)**

```
_cell_length_a      2.7954405934
_cell_length_b      2.7954405934
_cell_length_c      30.0000000000
_cell_angle_alpha   90.0000000000
_cell_angle_beta    90.0000000000
_cell_angle_gamma   90.0000000000
_cell_volume        234.4346433375

_symmetry_space_group_name_H-M "P 4/n 21/m 2/m (origin choice 2)"
_symmetry_Int_Tables_number 129
_space_group.reference_setting '129:-P 4a 2a'
_space_group.transform_Pp_abc a,b,c;0,0,0

loop_
_space_group_symop_id
_space_group_symop_operation_xyz
1 x,y,z
2 x+1/2,-y,-z
3 -x,y+1/2,-z
4 -x+1/2,-y+1/2,z
5 -y,-x,-z
6 -y+1/2,x,z
7 y,-x+1/2,z
8 y+1/2,x+1/2,-z
9 -x,-y,-z
10 -x+1/2,y,z
11 x,-y+1/2,z
12 x+1/2,y+1/2,-z
13 y,x,z
14 y+1/2,-x,-z
15 -y,x+1/2,-z
16 -y+1/2,-x+1/2,z

loop_
_atom_site_label
_atom_site_type_symbol
_atom_site_symmetry_multiplicity
_atom_site_Wyckoff_symbol
_atom_site_fract_x
_atom_site_fract_y
_atom_site_fract_z
_atom_site_occupancy
_atom_site_fract_symmform
Li1  Li   2  c   0.25000   0.25000   0.09822   1.00000  0,0,Dz
Li2  Li   2  c   0.25000   0.25000  -0.03343   1.00000  0,0,Dz
F1   F    2  c   0.25000   0.25000   0.89961   1.00000  0,0,Dz
F2   F    2  c   0.25000   0.25000   0.03301   1.00000  0,0,Dz

# end of cif
```

**#LiF(3D)**

```
_cell_length_a      3.9533500000
_cell_length_b      3.9533500000
_cell_length_c      3.9533500000
_cell_angle_alpha   90.0000000000
_cell_angle_beta    90.0000000000
_cell_angle_gamma   90.0000000000
_cell_volume        61.7868131492

_symmetry_space_group_name_H-M "F 4/m -3 2/m"
_symmetry_Int_Tables_number 225
_space_group.reference_setting '225:-F 4 2 3'
_space_group.transform_Pp_abc a,b,c;0,0,0

loop_
_space_group_symop_id
_space_group_symop_operation_xyz
1 x,y,z
2 x,-y,-z
3 -x,y,-z
4 -x,-y,z
5 y,z,x
6 y,-z,-x
7 -y,z,-x
8 -y,-z,x
9 z,x,y
10 z,-x,-y
11 -z,x,-y
12 -z,-x,y
13 -y,-x,-z
14 -y,x,z
15 y,-x,z
16 y,x,-z
17 -x,-z,-y
18 -x,z,y
19 x,-z,y
20 x,z,-y
21 -z,-y,-x
22 -z,y,x
23 z,-y,x
24 z,y,-x
25 -x,-y,-z
26 -x,y,z
27 x,-y,z
28 x,y,-z
29 -y,-z,-x
30 -y,z,x
31 y,-z,x
32 y,z,-x
33 -z,-x,-y
34 -z,x,y
35 z,-x,y
36 z,x,-y
37 y,x,z
38 y,-x,-z
39 -y,x,-z
40 -y,-x,z
41 x,z,y
42 x,-z,-y
43 -x,z,-y
44 -x,-z,y
45 z,y,x
46 z,-y,-x
47 -z,y,-x
```

```
48  -z,-y,x
49  x,y+1/2,z+1/2
50  x,-y+1/2,-z+1/2
51  -x,y+1/2,-z+1/2
52  -x,-y+1/2,z+1/2
53  y,z+1/2,x+1/2
54  y,-z+1/2,-x+1/2
55  -y,z+1/2,-x+1/2
56  -y,-z+1/2,x+1/2
57  z,x+1/2,y+1/2
58  z,-x+1/2,-y+1/2
59  -z,x+1/2,-y+1/2
60  -z,-x+1/2,y+1/2
61  -y,-x+1/2,-z+1/2
62  -y,x+1/2,z+1/2
63  y,-x+1/2,z+1/2
64  y,x+1/2,-z+1/2
65  -x,-z+1/2,-y+1/2
66  -x,z+1/2,y+1/2
67  x,-z+1/2,y+1/2
68  x,z+1/2,-y+1/2
69  -z,-y+1/2,-x+1/2
70  -z,y+1/2,x+1/2
71  z,-y+1/2,x+1/2
72  z,y+1/2,-x+1/2
73  -x,-y+1/2,-z+1/2
74  -x,y+1/2,z+1/2
75  x,-y+1/2,z+1/2
76  x,y+1/2,-z+1/2
77  -y,-z+1/2,-x+1/2
78  -y,z+1/2,x+1/2
79  y,-z+1/2,x+1/2
80  y,z+1/2,-x+1/2
81  -z,-x+1/2,-y+1/2
82  -z,x+1/2,y+1/2
83  z,-x+1/2,y+1/2
84  z,x+1/2,-y+1/2
85  y,x+1/2,z+1/2
86  y,-x+1/2,-z+1/2
87  -y,x+1/2,-z+1/2
88  -y,-x+1/2,z+1/2
89  x,z+1/2,y+1/2
90  x,-z+1/2,-y+1/2
91  -x,z+1/2,-y+1/2
92  -x,-z+1/2,y+1/2
93  z,y+1/2,x+1/2
94  z,-y+1/2,-x+1/2
95  -z,y+1/2,-x+1/2
96  -z,-y+1/2,x+1/2
97  x+1/2,y,z+1/2
98  x+1/2,-y,-z+1/2
99  -x+1/2,y,-z+1/2
100 -x+1/2,-y,z+1/2
101 y+1/2,z,x+1/2
102 y+1/2,-z,-x+1/2
103 -y+1/2,z,-x+1/2
104 -y+1/2,-z,x+1/2
105 z+1/2,x,y+1/2
106 z+1/2,-x,-y+1/2
107 -z+1/2,x,-y+1/2
108 -z+1/2,-x,y+1/2
109 -y+1/2,-x,-z+1/2
110 -y+1/2,x,z+1/2
111 y+1/2,-x,z+1/2
112 y+1/2,x,-z+1/2
```

```
113 -x+1/2,-z,-y+1/2
114 -x+1/2,z,y+1/2
115 x+1/2,-z,y+1/2
116 x+1/2,z,-y+1/2
117 -z+1/2,-y,-x+1/2
118 -z+1/2,y,x+1/2
119 z+1/2,-y,x+1/2
120 z+1/2,y,-x+1/2
121 -x+1/2,-y,-z+1/2
122 -x+1/2,y,z+1/2
123 x+1/2,-y,z+1/2
124 x+1/2,y,-z+1/2
125 -y+1/2,-z,-x+1/2
126 -y+1/2,z,x+1/2
127 y+1/2,-z,x+1/2
128 y+1/2,z,-x+1/2
129 -z+1/2,-x,-y+1/2
130 -z+1/2,x,y+1/2
131 z+1/2,-x,y+1/2
132 z+1/2,x,-y+1/2
133 y+1/2,x,z+1/2
134 y+1/2,-x,-z+1/2
135 -y+1/2,x,-z+1/2
136 -y+1/2,-x,z+1/2
137 x+1/2,z,y+1/2
138 x+1/2,-z,-y+1/2
139 -x+1/2,z,-y+1/2
140 -x+1/2,-z,y+1/2
141 z+1/2,y,x+1/2
142 z+1/2,-y,-x+1/2
143 -z+1/2,y,-x+1/2
144 -z+1/2,-y,x+1/2
145 x+1/2,y+1/2,z
146 x+1/2,-y+1/2,-z
147 -x+1/2,y+1/2,-z
148 -x+1/2,-y+1/2,z
149 y+1/2,z+1/2,x
150 y+1/2,-z+1/2,-x
151 -y+1/2,z+1/2,-x
152 -y+1/2,-z+1/2,x
153 z+1/2,x+1/2,y
154 z+1/2,-x+1/2,-y
155 -z+1/2,x+1/2,-y
156 -z+1/2,-x+1/2,y
157 -y+1/2,-x+1/2,-z
158 -y+1/2,x+1/2,z
159 y+1/2,-x+1/2,z
160 y+1/2,x+1/2,-z
161 -x+1/2,-z+1/2,-y
162 -x+1/2,z+1/2,y
163 x+1/2,-z+1/2,y
164 x+1/2,z+1/2,-y
165 -z+1/2,-y+1/2,-x
166 -z+1/2,y+1/2,x
167 z+1/2,-y+1/2,x
168 z+1/2,y+1/2,-x
169 -x+1/2,-y+1/2,-z
170 -x+1/2,y+1/2,z
171 x+1/2,-y+1/2,z
172 x+1/2,y+1/2,-z
173 -y+1/2,-z+1/2,-x
174 -y+1/2,z+1/2,x
175 y+1/2,-z+1/2,x
176 y+1/2,z+1/2,-x
177 -z+1/2,-x+1/2,-y
```

```
178 -z+1/2,x+1/2,y
179 z+1/2,-x+1/2,y
180 z+1/2,x+1/2,-y
181 y+1/2,x+1/2,z
182 y+1/2,-x+1/2,-z
183 -y+1/2,x+1/2,-z
184 -y+1/2,-x+1/2,z
185 x+1/2,z+1/2,y
186 x+1/2,-z+1/2,-y
187 -x+1/2,z+1/2,-y
188 -x+1/2,-z+1/2,y
189 z+1/2,y+1/2,x
190 z+1/2,-y+1/2,-x
191 -z+1/2,y+1/2,-x
192 -z+1/2,-y+1/2,x

loop_
_atom_site_label
_atom_site_type_symbol
_atom_site_symmetry_multiplicity
_atom_site_Wyckoff_symbol
_atom_site_fract_x
_atom_site_fract_y
_atom_site_fract_z
_atom_site_occupancy
_atom_site_fract_symmform
Li1  Li   4  b   0.50000   0.50000   0.50000   1.00000 0,0,0
F1   F    4  a   0.00000   0.00000   0.00000   1.00000 0,0,0

# end of cif
```

**#Li|(LiF)₃|F**

```
_cell_length_a     2.7954405934
_cell_length_b     2.7954405934
_cell_length_c     30.0000000000
_cell_angle_alpha  90.0000000000
_cell_angle_beta   90.0000000000
_cell_angle_gamma  90.0000000000
_cell_volume       234.4346433375

_symmetry_space_group_name_H-M "P 4 m m"
_symmetry_Int_Tables_number 99
_space_group.reference_setting '099:P 4 -2'
_space_group.transform_Pp_abc a,b,c;0,0,0

loop_
_space_group_symop_id
_space_group_symop_operation_xyz
1 x,y,z
2 -x,-y,z
3 -y,x,z
4 y,-x,z
5 -x,y,z
6 x,-y,z
7 y,x,z
8 -y,-x,z

loop_
_atom_site_label
_atom_site_type_symbol
_atom_site_symmetry_multiplicity
_atom_site_Wyckoff_symbol
_atom_site_fract_x
_atom_site_fract_y
_atom_site_fract_z
_atom_site_occupancy
_atom_site_fract_symmform
Li1 Li   1 b   0.50000   0.50000   0.45947   1.00000 0,0,Dz
Li2 Li   1 a   0.00000   0.00000   0.37922   1.00000 0,0,Dz
Li3 Li   1 b   0.50000   0.50000   0.29886   1.00000 0,0,Dz
Li4 Li   1 a   0.00000   0.00000   0.21789   1.00000 0,0,Dz
F1  F    1 b   0.50000   0.50000   0.40129   1.00000 0,0,Dz
F2  F    1 a   0.00000   0.00000   0.32090   1.00000 0,0,Dz
F3  F    1 b   0.50000   0.50000   0.24045   1.00000 0,0,Dz
F4  F    1 a   0.00000   0.00000   0.15828   1.00000 0,0,Dz

# end of cif
```

**#Tl|(LiF)₃|F**

```
_cell_length_a      3.9533500000
_cell_length_b      3.9533500000
_cell_length_c      30.0000000000
_cell_angle_alpha   90.0000000000
_cell_angle_beta    90.0000000000
_cell_angle_gamma   90.0000000000
_cell_volume        468.8692866750

_symmetry_space_group_name_H-M "P 4 m m"
_symmetry_Int_Tables_number 99
_space_group.reference_setting '099:P 4 -2'
_space_group.transform_Pp_abc a,b,c;0,0,0

loop_
_space_group_symop_id
_space_group_symop_operation_xyz
1 x,y,z
2 -x,-y,z
3 -y,x,z
4 y,-x,z
5 -x,y,z
6 x,-y,z
7 y,x,z
8 -y,-x,z

loop_
_atom_type_symbol
Tl
Li
F

loop_
_atom_site_label
_atom_site_type_symbol
_atom_site_symmetry_multiplicity
_atom_site_Wyckoff_symbol
_atom_site_fract_x
_atom_site_fract_y
_atom_site_fract_z
_atom_site_occupancy
_atom_site_fract_symmform
Tl1  Tl   1 b   0.5000000000   0.5000000000   0.1478700000   1.0000000000  0,0,Dz
Li1  Li   2 c   0.5000000000   0.0000000000   0.3987000000   1.0000000000  0,0,Dz
Li2  Li   1 b   0.5000000000   0.5000000000   0.3186400000   1.0000000000  0,0,Dz
Li3  Li   1 a   0.0000000000   0.0000000000   0.3193800000   1.0000000000  0,0,Dz
Li4  Li   2 c   0.5000000000   0.0000000000   0.2398400000   1.0000000000  0,0,Dz
F1   F    2 c   0.5000000000   0.0000000000   0.4592200000   1.0000000000  0,0,Dz
F2   F    1 b   0.5000000000   0.5000000000   0.3778500000   1.0000000000  0,0,Dz
F3   F    1 a   0.0000000000   0.0000000000   0.3783900000   1.0000000000  0,0,Dz
F4   F    2 c   0.5000000000   0.0000000000   0.2989700000   1.0000000000  0,0,Dz
F5   F    1 b   0.5000000000   0.5000000000   0.2249600000   1.0000000000  0,0,Dz
F6   F    1 a   0.0000000000   0.0000000000   0.2134700000   1.0000000000  0,0,Dz

# end of cif
```

# #(Tl₄)|(Li₈F₈)₃|F(PtF₅)

```
_cell_length_a        7.9067000000
_cell_length_b        7.9067000000
_cell_length_c        30.0000000000
_cell_angle_alpha     90.0000000000
_cell_angle_beta      90.0000000000
_cell_angle_gamma     90.0000000000
_cell_volume          1875.4771467000

_symmetry_space_group_name_H-M "P 1"
_symmetry_Int_Tables_number 1
_space_group.reference_setting '001:P 1'
_space_group.transform_Pp_abc a,b,c;0,0,0

loop_
_space_group_symop_id
_space_group_symop_operation_xyz
1 x,y,z

loop_
_atom_site_label
_atom_site_type_symbol
_atom_site_symmetry_multiplicity
_atom_site_Wyckoff_symbol
_atom_site_fract_x
_atom_site_fract_y
_atom_site_fract_z
_atom_site_occupancy
_atom_site_fract_symmform
Pt1   Pt   1 a   0.03275   0.21469   0.13832   1.00000 Dx,Dy,Dz
Li1   Li   1 a   0.01471  -0.01067   0.23482   1.00000 Dx,Dy,Dz
Li2   Li   1 a   0.00932   0.47006   0.22790   1.00000 Dx,Dy,Dz
Li3   Li   1 a   0.51220  -0.00942   0.24295   1.00000 Dx,Dy,Dz
Li4   Li   1 a   0.51081   0.48499   0.24224   1.00000 Dx,Dy,Dz
Li5   Li   1 a   0.26027   0.23470   0.23461   1.00000 Dx,Dy,Dz
Li6   Li   1 a   0.25855   0.73729   0.24278   1.00000 Dx,Dy,Dz
Li7   Li   1 a   0.77730   0.23943   0.22838   1.00000 Dx,Dy,Dz
Li8   Li   1 a   0.76491   0.73878   0.24220   1.00000 Dx,Dy,Dz
Li9   Li   1 a   0.01164   0.23805   0.31435   1.00000 Dx,Dy,Dz
Li10  Li   1 a   0.01120   0.73508   0.31017   1.00000 Dx,Dy,Dz
Li11  Li   1 a   0.51436   0.23850   0.31021   1.00000 Dx,Dy,Dz
Li12  Li   1 a   0.51257   0.73695   0.31000   1.00000 Dx,Dy,Dz
Li13  Li   1 a   0.25630  -0.00673   0.30857   1.00000 Dx,Dy,Dz
Li14  Li   1 a   0.25357   0.48149   0.30773   1.00000 Dx,Dy,Dz
Li15  Li   1 a   0.76815  -0.00420   0.30781   1.00000 Dx,Dy,Dz
Li16  Li   1 a   0.76902   0.48079   0.30818   1.00000 Dx,Dy,Dz
Li17  Li   1 a   0.01180  -0.01284   0.38360   1.00000 Dx,Dy,Dz
Li18  Li   1 a   0.01196   0.48742   0.38425   1.00000 Dx,Dy,Dz
Li19  Li   1 a   0.51244  -0.00893   0.37938   1.00000 Dx,Dy,Dz
Li20  Li   1 a   0.51177   0.48400   0.37954   1.00000 Dx,Dy,Dz
Li21  Li   1 a   0.26246   0.23787   0.38360   1.00000 Dx,Dy,Dz
Li22  Li   1 a   0.25852   0.73716   0.37939   1.00000 Dx,Dy,Dz
Li23  Li   1 a   0.76221   0.23772   0.38420   1.00000 Dx,Dy,Dz
Li24  Li   1 a   0.76565   0.73783   0.37956   1.00000 Dx,Dy,Dz
F1    F    1 a  -0.01457   0.01505   0.17537   1.00000 Dx,Dy,Dz
F2    F    1 a   0.23124   0.26708   0.17534   1.00000 Dx,Dy,Dz
F3    F    1 a   0.16987   0.08011   0.10096   1.00000 Dx,Dy,Dz
F4    F    1 a   0.89174   0.35374   0.17932   1.00000 Dx,Dy,Dz
F5    F    1 a   0.83613   0.16538   0.10484   1.00000 Dx,Dy,Dz
F6    F    1 a   0.07743   0.41204   0.10471   1.00000 Dx,Dy,Dz
F7    F    1 a   0.26575  -0.01594   0.24742   1.00000 Dx,Dy,Dz
F8    F    1 a   0.26213   0.49216   0.24675   1.00000 Dx,Dy,Dz
F9    F    1 a   0.75736  -0.01242   0.24680   1.00000 Dx,Dy,Dz
F10   F    1 a   0.75678   0.49293   0.24778   1.00000 Dx,Dy,Dz
F11   F    1 a   0.01554   0.23392   0.25353   1.00000 Dx,Dy,Dz
F12   F    1 a   0.01132   0.73630   0.24715   1.00000 Dx,Dy,Dz
F13   F    1 a   0.51355   0.23823   0.24711   1.00000 Dx,Dy,Dz
F14   F    1 a   0.51141   0.73822   0.24680   1.00000 Dx,Dy,Dz
F15   F    1 a   0.26461   0.23763   0.32185   1.00000 Dx,Dy,Dz
F16   F    1 a   0.26225   0.73737   0.31787   1.00000 Dx,Dy,Dz
```

```
F17   F    1 a   0.76078   0.23777   0.32271   1.00000 Dx,Dy,Dz
F18   F    1 a   0.76145   0.73848   0.31808   1.00000 Dx,Dy,Dz
F19   F    1 a   0.01201  -0.01500   0.32185   1.00000 Dx,Dy,Dz
F20   F    1 a   0.01185   0.48881   0.32277   1.00000 Dx,Dy,Dz
F21   F    1 a   0.51222  -0.01265   0.31784   1.00000 Dx,Dy,Dz
F22   F    1 a   0.51114   0.48817   0.31805   1.00000 Dx,Dy,Dz
F23   F    1 a   0.01264   0.23695   0.39183   1.00000 Dx,Dy,Dz
F24   F    1 a   0.01216   0.73772   0.39078   1.00000 Dx,Dy,Dz
F25   F    1 a   0.51193   0.23747   0.39071   1.00000 Dx,Dy,Dz
F26   F    1 a   0.51202   0.73761   0.39038   1.00000 Dx,Dy,Dz
F27   F    1 a   0.26528  -0.01562   0.39617   1.00000 Dx,Dy,Dz
F28   F    1 a   0.26510   0.49068   0.39689   1.00000 Dx,Dy,Dz
F29   F    1 a   0.75892  -0.01545   0.39684   1.00000 Dx,Dy,Dz
F30   F    1 a   0.75884   0.49077   0.39704   1.00000 Dx,Dy,Dz
Tl1   Tl   1 a   0.30333  -0.05353   0.47689   1.00000 Dx,Dy,Dz
Tl2   Tl   1 a   0.30357   0.53225   0.47734   1.00000 Dx,Dy,Dz
Tl3   Tl   1 a   0.71756  -0.05378   0.47732   1.00000 Dx,Dy,Dz
Tl4   Tl   1 a   0.71718   0.53256   0.47741   1.00000 Dx,Dy,Dz

# end of cif
```

# (Tl)₃|(Li₂F₂)₃|(KrF₂)

```
_cell_length_a       3.9533500000
_cell_length_b       3.9533500000
_cell_length_c       30.0000000000
_cell_angle_alpha    90.0000000000
_cell_angle_beta     90.0000000000
_cell_angle_gamma    90.0000000000
_cell_volume         468.8692866750

_symmetry_space_group_name_H-M "P m m 2"
_symmetry_Int_Tables_number 25
_space_group.reference_setting '025:P 2 -2'
_space_group.transform_Pp_abc a,b,c;0,0,0

loop_
_space_group_symop_id
_space_group_symop_operation_xyz
1 x,y,z
2 -x,-y,z
3 -x,y,z
4 x,-y,z

loop_
_atom_type_symbol
Kr
Tl
Li
F

loop_
_atom_site_label
_atom_site_type_symbol
_atom_site_symmetry_multiplicity
_atom_site_Wyckoff_symbol
_atom_site_fract_x
_atom_site_fract_y
_atom_site_fract_z
_atom_site_occupancy
_atom_site_fract_symmform
Kr1  Kr    1 d   0.5000000000   0.5000000000   0.0965100000   1.0000000000  0,0,Dz
Tl1  Tl    1 c   0.5000000000   0.0000000000   0.5937900000   1.0000000000  0,0,Dz
Tl2  Tl    1 b   0.0000000000   0.5000000000   0.5313600000   1.0000000000  0,0,Dz
Tl3  Tl    1 c   0.5000000000   0.0000000000   0.4684000000   1.0000000000  0,0,Dz
Li1  Li    1 d   0.5000000000   0.5000000000   0.3718600000   1.0000000000  0,0,Dz
Li2  Li    1 a   0.0000000000   0.0000000000   0.3713200000   1.0000000000  0,0,Dz
Li3  Li    1 c   0.5000000000   0.0000000000   0.3019800000   1.0000000000  0,0,Dz
Li4  Li    1 b   0.0000000000   0.5000000000   0.3033500000   1.0000000000  0,0,Dz
Li5  Li    1 d   0.5000000000   0.5000000000   0.2288100000   1.0000000000  0,0,Dz
Li6  Li    1 a   0.0000000000   0.0000000000   0.2375700000   1.0000000000  0,0,Dz
F1   F     1 d   0.5000000000   0.5000000000   0.0310000000   1.0000000000  0,0,Dz
F2   F     1 c   0.5000000000   0.0000000000   0.3810100000   1.0000000000  0,0,Dz
F3   F     1 b   0.0000000000   0.5000000000   0.3800400000   1.0000000000  0,0,Dz
F4   F     1 c   0.5000000000   0.0000000000   0.2397800000   1.0000000000  0,0,Dz
F5   F     1 b   0.0000000000   0.5000000000   0.2404200000   1.0000000000  0,0,Dz
F6   F     1 a   0.0000000000   0.0000000000   0.3088000000   1.0000000000  0,0,Dz
F7   F     1 d   0.5000000000   0.5000000000   0.3102900000   1.0000000000  0,0,Dz
F8   F     1 d   0.5000000000   0.5000000000   0.1649300000   1.0000000000  0,0,Dz

# end of cif
```

# #(Tl$_2$)|(Li$_4$F$_4$)$_3$|(PbO$_2$)$_3$

```
_cell_length_a      7.9067000000
_cell_length_b      7.9067000000
_cell_length_c      40.0000000000
_cell_angle_alpha   90.0000000000
_cell_angle_beta    90.0000000000
_cell_angle_gamma   90.0000000000
_cell_volume        2500.6361956000

_symmetry_space_group_name_H-M "C m m 2"
_symmetry_Int_Tables_number 35
_space_group.reference_setting '035:C 2 -2'
_space_group.transform_Pp_abc a,b,c;0,0,0

loop_
_space_group_symop_id
_space_group_symop_operation_xyz
1 x,y,z
2 -x,-y,z
3 -x,y,z
4 x,-y,z
5 x+1/2,y+1/2,z
6 -x+1/2,-y+1/2,z
7 -x+1/2,y+1/2,z
8 x+1/2,-y+1/2,z

loop_
_atom_site_label
_atom_site_type_symbol
_atom_site_symmetry_multiplicity
_atom_site_Wyckoff_symbol
_atom_site_fract_x
_atom_site_fract_y
_atom_site_fract_z
_atom_site_occupancy
_atom_site_fract_symmform
Pb1  Pb   2 a   0.00000   0.00000   0.12693   1.00000  0,0,Dz
Pb2  Pb   2 b   0.00000   0.50000   0.15544   1.00000  0,0,Dz
Pb3  Pb   2 a   0.00000   0.00000   0.20308   1.00000  0,0,Dz
O1   O    4 e   0.00000   0.73853   0.11961   1.00000  0,Dy,Dz
O2   O    4 d   0.79725   0.00000   0.16418   1.00000  Dx,0,Dz
O3   O    4 e   0.00000   0.75321   0.21413   1.00000  0,Dy,Dz
Li1  Li   4 e   0.00000   0.73763   0.26096   1.00000  0,Dy,Dz
Li2  Li   4 e   0.00000   0.75046   0.37595   1.00000  0,Dy,Dz
Li3  Li   4 d   0.25277   0.00000   0.27381   1.00000  Dx,0,Dz
Li4  Li   2 a   0.00000   0.00000   0.31948   1.00000  0,0,Dz
Li5  Li   2 b   0.00000   0.50000   0.32093   1.00000  0,0,Dz
Li6  Li   4 c   0.25000   0.25000   0.32322   1.00000  0,0,Dz
Li7  Li   4 d   0.24931   0.00000   0.37490   1.00000  Dx,0,Dz
F1   F    4 c   0.25000   0.25000   0.27677   1.00000  0,0,Dz
F2   F    2 a   0.00000   0.00000   0.27308   1.00000  0,0,Dz
F3   F    2 b   0.00000   0.50000   0.27503   1.00000  0,0,Dz
F4   F    4 d   0.24921   0.00000   0.32809   1.00000  Dx,0,Dz
F5   F    4 e   0.00000   0.75035   0.33033   1.00000  0,Dy,Dz
F6   F    2 a   0.00000   0.00000   0.38761   1.00000  0,0,Dz
F7   F    2 b   0.00000   0.50000   0.38711   1.00000  0,0,Dz
F8   F    4 c   0.25000   0.25000   0.38154   1.00000  0,0,Dz
Tl1  Tl   2 a   0.00000   0.00000   0.44935   1.00000  0,0,Dz
Tl2  Tl   2 b   0.00000   0.50000   0.44905   1.00000  0,0,Dz

# end of cif
```

# #(Tl₄)₃|(Li₈F₈)₃|(Pb₂O₄)₅

```
_cell_length_a        7.9067000000
_cell_length_b        7.9067000000
_cell_length_c        40.0000000000
_cell_angle_alpha     90.0000000000
_cell_angle_beta      90.0000000000
_cell_angle_gamma     90.0000000000
_cell_volume          2500.6361956000

_symmetry_space_group_name_H-M "C m m 2"
_symmetry_Int_Tables_number 35
_space_group.reference_setting '035:C 2 -2'
_space_group.transform_Pp_abc a,b,c;0,0,0

loop_
_space_group_symop_id
_space_group_symop_operation_xyz
1 x,y,z
2 -x,-y,z
3 -x,y,z
4 x,-y,z
5 x+1/2,y+1/2,z
6 -x+1/2,-y+1/2,z
7 -x+1/2,y+1/2,z
8 x+1/2,-y+1/2,z

loop_
_atom_site_label
_atom_site_type_symbol
_atom_site_symmetry_multiplicity
_atom_site_Wyckoff_symbol
_atom_site_fract_x
_atom_site_fract_y
_atom_site_fract_z
_atom_site_occupancy
_atom_site_fract_symmform
Pb1  Pb   2  a   0.00000   0.00000   0.05533   1.00000  0,0,Dz
Pb2  Pb   2  b   0.00000   0.50000   0.09114   1.00000  0,0,Dz
Pb3  Pb   2  a   0.00000   0.00000   0.13178   1.00000  0,0,Dz
Pb4  Pb   2  b   0.00000   0.50000   0.17500   1.00000  0,0,Dz
Pb5  Pb   2  a   0.00000   0.00000   0.20717   1.00000  0,0,Dz
O1   O    4  e   0.00000   0.73558   0.04858   1.00000  0,Dy,Dz
O2   O    4  d   0.80577   0.00000   0.09347   1.00000  Dx,0,Dz
O3   O    4  e   0.00000   0.68308   0.13311   1.00000  0,Dy,Dz
O4   O    4  d   0.80345   0.00000   0.17127   1.00000  Dx,0,Dz
O5   O    4  e   0.00000   0.72550   0.21402   1.00000  0,Dy,Dz
Li1  Li   4  e   0.00000   0.72786   0.26173   1.00000  0,Dy,Dz
Li2  Li   4  e   0.00000   0.75040   0.37198   1.00000  0,Dy,Dz
Li3  Li   4  d   0.25674   0.00000   0.27468   1.00000  Dx,0,Dz
Li4  Li   2  a   0.00000   0.00000   0.32293   1.00000  0,0,Dz
Li5  Li   2  b   0.00000   0.50000   0.32298   1.00000  0,0,Dz
Li6  Li   4  c   0.25000   0.25000   0.32066   1.00000  0,0,Dz
Li7  Li   4  d   0.24943   0.00000   0.37172   1.00000  Dx,0,Dz
F1   F    4  c   0.25000   0.25000   0.27400   1.00000  0,0,Dz
F2   F    2  a   0.00000   0.00000   0.26696   1.00000  0,0,Dz
F3   F    2  b   0.00000   0.50000   0.27559   1.00000  0,0,Dz
F4   F    4  d   0.24684   0.00000   0.32421   1.00000  Dx,0,Dz
F5   F    4  e   0.00000   0.75243   0.32567   1.00000  0,Dy,Dz
F6   F    2  a   0.00000   0.00000   0.37655   1.00000  0,0,Dz
F7   F    2  b   0.00000   0.50000   0.37564   1.00000  0,0,Dz
F8   F    4  c   0.25000   0.25000   0.37727   1.00000  0,0,Dz
Tl1  Tl   4  c   0.25000   0.25000   0.44587   1.00000  0,0,Dz
Tl2  Tl   2  a   0.00000   0.00000   0.49221   1.00000  0,0,Dz
Tl3  Tl   2  b   0.00000   0.50000   0.49217   1.00000  0,0,Dz
Tl4  Tl   4  c   0.25000   0.25000   0.53879   1.00000  0,0,Dz

# end of cif
```

# #(Tl$_2$)$_3$|(Sr$_4$Ti$_2$O$_8$)|(KrF$_2$)

```
_cell_length_a      3.8821100000
_cell_length_b      3.8821100000
_cell_length_c      30.0000000000
_cell_angle_alpha   90.0000000000
_cell_angle_beta    90.0000000000
_cell_angle_gamma   90.0000000000
_cell_volume        452.1233415630

_symmetry_space_group_name_H-M "P 4 m m"
_symmetry_Int_Tables_number 99
_space_group.reference_setting '099:P 4 -2'
_space_group.transform_Pp_abc a,b,c;0,0,0

loop_
_space_group_symop_id
_space_group_symop_operation_xyz
1 x,y,z
2 -x,-y,z
3 -y,x,z
4 y,-x,z
5 -x,y,z
6 x,-y,z
7 y,x,z
8 -y,-x,z

loop_
_atom_site_label
_atom_site_type_symbol
_atom_site_symmetry_multiplicity
_atom_site_Wyckoff_symbol
_atom_site_fract_x
_atom_site_fract_y
_atom_site_fract_z
_atom_site_occupancy
_atom_site_fract_symmform
Kr1  Kr   1 a   0.00000   0.00000   0.18796   1.00000  0,0,Dz
F1   F    1 a   0.00000   0.00000   0.26507   1.00000  0,0,Dz
F2   F    1 a   0.00000   0.00000   0.12019   1.00000  0,0,Dz
Tl1  Tl   1 b   0.50000   0.50000   0.55628   1.00000  0,0,Dz
Sr1  Sr   1 a   0.00000   0.00000   0.46563   1.00000  0,0,Dz
Sr2  Sr   1 a   0.00000   0.00000   0.33850   1.00000  0,0,Dz
Ti1  Ti   1 b   0.50000   0.50000   0.40441   1.00000  0,0,Dz
O1   O    2 c   0.50000   0.00000   0.41103   1.00000  0,0,Dz
O2   O    1 b   0.50000   0.50000   0.48134   1.00000  0,0,Dz
O3   O    1 b   0.50000   0.50000   0.34486   1.00000  0,0,Dz

# end of cif
```

# #(Tl$_2$)$_3$|(Sr$_4$Ti$_2$O$_8$)|(OsO$_4$)

```
_cell_length_a      7.7642200000
_cell_length_b      7.7642200000
_cell_length_c      30.0000000000
_cell_angle_alpha   90.0000000000
_cell_angle_beta    90.0000000000
_cell_angle_gamma   90.0000000000
_cell_volume        1808.4933662520

_symmetry_space_group_name_H-M "P 4 m m"
_symmetry_Int_Tables_number 99
_space_group.reference_setting '099:P 4 -2'
_space_group.transform_Pp_abc a,b,c;0,0,0

loop_
_space_group_symop_id
_space_group_symop_operation_xyz
1 x,y,z
2 -x,-y,z
3 -y,x,z
4 y,-x,z
5 -x,y,z
6 x,-y,z
7 y,x,z
8 -y,-x,z

loop_
_atom_site_label
_atom_site_type_symbol
_atom_site_symmetry_multiplicity
_atom_site_Wyckoff_symbol
_atom_site_fract_x
_atom_site_fract_y
_atom_site_fract_z
_atom_site_occupancy
_atom_site_fract_symmform
Tl1  Tl   4 d   0.20749   0.20749   0.57200   1.00000  Dx,Dx,Dz
Tl2  Tl   2 c   0.50000   0.00000   0.51126   1.00000  0,0,Dz
Tl3  Tl   1 b   0.50000   0.50000   0.53759   1.00000  0,0,Dz
Tl4  Tl   1 a   0.00000   0.00000   0.48825   1.00000  0,0,Dz
Tl5  Tl   4 d   0.27883   0.27883   0.45665   1.00000  Dx,Dx,Dz
Sr1  Sr   2 c   0.50000   0.00000   0.37084   1.00000  0,0,Dz
Sr2  Sr   1 b   0.50000   0.50000   0.36679   1.00000  0,0,Dz
Sr3  Sr   1 a   0.00000   0.00000   0.37147   1.00000  0,0,Dz
Sr4  Sr   2 c   0.50000   0.00000   0.24878   1.00000  0,0,Dz
Sr5  Sr   1 b   0.50000   0.50000   0.21669   1.00000  0,0,Dz
Sr6  Sr   1 a   0.00000   0.00000   0.25100   1.00000  0,0,Dz
Ti1  Ti   4 d   0.25157   0.25157   0.30820   1.00000  Dx,Dx,Dz
O1   O    4 e   0.75218   0.00000   0.31459   1.00000  Dx,0,Dz
O2   O    4 f   0.75581   0.50000   0.31259   1.00000  Dx,0,Dz
O3   O    4 d   0.25494   0.25494   0.38166   1.00000  Dx,Dx,Dz
O4   O    4 d   0.24131   0.24131   0.24789   1.00000  Dx,Dx,Dz
O5   O    4 f   0.82085   0.50000   0.17507   1.00000  Dx,0,Dz
O6   O    4 e   0.31056   0.00000   0.10728   1.00000  Dx,0,Dz
Os1  Os   2 c   0.50000   0.00000   0.13720   1.00000  0,0,Dz

# end of cif
```

**#(Tl$_2$)$_3$|(Sr$_4$Ti$_2$O$_8$)|(RuO$_4$)**

```
_cell_length_a        7.7642200000
_cell_length_b        7.7642200000
_cell_length_c       30.0000000000
_cell_angle_alpha    90.0000000000
_cell_angle_beta     90.0000000000
_cell_angle_gamma    90.0000000000
_cell_volume       1808.4933662520

_symmetry_space_group_name_H-M "P 4 m m"
_symmetry_Int_Tables_number 99
_space_group.reference_setting '099:P 4 -2'
_space_group.transform_Pp_abc a,b,c;0,0,0

loop_
_space_group_symop_id
_space_group_symop_operation_xyz
1 x,y,z
2 -x,-y,z
3 -y,x,z
4 y,-x,z
5 -x,y,z
6 x,-y,z
7 y,x,z
8 -y,-x,z

loop_
_atom_site_label
_atom_site_type_symbol
_atom_site_symmetry_multiplicity
_atom_site_Wyckoff_symbol
_atom_site_fract_x
_atom_site_fract_y
_atom_site_fract_z
_atom_site_occupancy
_atom_site_fract_symmform
Tl1  Tl   4 d   0.20880   0.20880   0.57265   1.00000 Dx,Dx,Dz
Tl2  Tl   2 c   0.50000   0.00000   0.51162   1.00000 0,0,Dz
Tl3  Tl   1 b   0.50000   0.50000   0.53801   1.00000 0,0,Dz
Tl4  Tl   1 a   0.00000   0.00000   0.48899   1.00000 0,0,Dz
Tl5  Tl   4 d   0.27887   0.27887   0.45712   1.00000 Dx,Dx,Dz
Sr1  Sr   2 c   0.50000   0.00000   0.37050   1.00000 0,0,Dz
Sr2  Sr   1 b   0.50000   0.50000   0.36575   1.00000 0,0,Dz
Sr3  Sr   1 a   0.00000   0.00000   0.37121   1.00000 0,0,Dz
Sr4  Sr   2 c   0.50000   0.00000   0.24751   1.00000 0,0,Dz
Sr5  Sr   1 b   0.50000   0.50000   0.20095   1.00000 0,0,Dz
Sr6  Sr   1 a   0.00000   0.00000   0.24965   1.00000 0,0,Dz
Ti1  Ti   4 d   0.25120   0.25120   0.30728   1.00000 Dx,Dx,Dz
O1   O    4 e   0.75246   0.00000   0.31600   1.00000 Dx,0,Dz
O2   O    4 f   0.75729   0.50000   0.31262   1.00000 Dx,0,Dz
O3   O    4 d   0.25713   0.25713   0.38366   1.00000 Dx,Dx,Dz
O4   O    4 d   0.23502   0.23502   0.24796   1.00000 Dx,Dx,Dz
O5   O    4 f   0.82172   0.50000   0.17504   1.00000 Dx,0,Dz
O6   O    4 e   0.31217   0.00000   0.10851   1.00000 Dx,0,Dz
Ru1  Ru   2 c   0.50000   0.00000   0.13726   1.00000 0,0,Dz

# end of cif
```

**#(Tl$_2$)$_3$|(Sr$_4$Ti$_2$O$_8$)|(XeO$_4$)**

```
_cell_length_a      7.7642200000
_cell_length_b      7.7642200000
_cell_length_c      30.0000000000
_cell_angle_alpha   90.0000000000
_cell_angle_beta    90.0000000000
_cell_angle_gamma   90.0000000000
_cell_volume        1808.4933662520

_symmetry_space_group_name_H-M "P 4 m m"
_symmetry_Int_Tables_number 99
_space_group.reference_setting '099:P 4 -2'
_space_group.transform_Pp_abc a,b,c;0,0,0

loop_
_space_group_symop_id
_space_group_symop_operation_xyz
1 x,y,z
2 -x,-y,z
3 -y,x,z
4 y,-x,z
5 -x,y,z
6 x,-y,z
7 y,x,z
8 -y,-x,z

loop_
_atom_site_label
_atom_site_type_symbol
_atom_site_symmetry_multiplicity
_atom_site_Wyckoff_symbol
_atom_site_fract_x
_atom_site_fract_y
_atom_site_fract_z
_atom_site_occupancy
_atom_site_fract_symmform
Tl1  Tl   4 d   0.21771   0.21771   0.57795   1.00000 Dx,Dx,Dz
Tl2  Tl   2 c   0.50000   0.00000   0.51631   1.00000 0,0,Dz
Tl3  Tl   1 b   0.50000   0.50000   0.53774   1.00000 0,0,Dz
Tl4  Tl   1 a   0.00000   0.00000   0.49865   1.00000 0,0,Dz
Tl5  Tl   4 d   0.27408   0.27408   0.46172   1.00000 Dx,Dx,Dz
Sr1  Sr   2 c   0.50000   0.00000   0.37150   1.00000 0,0,Dz
Sr2  Sr   1 b   0.50000   0.50000   0.36739   1.00000 0,0,Dz
Sr3  Sr   1 a   0.00000   0.00000   0.37365   1.00000 0,0,Dz
Sr4  Sr   2 c   0.50000   0.00000   0.24088   1.00000 0,0,Dz
Sr5  Sr   1 b   0.50000   0.50000   0.19065   1.00000 0,0,Dz
Sr6  Sr   1 a   0.00000   0.00000   0.25095   1.00000 0,0,Dz
Ti1  Ti   4 d   0.25048   0.25048   0.30801   1.00000 Dx,Dx,Dz
O1   O    4 e   0.75394   0.00000   0.32031   1.00000 Dx,0,Dz
O2   O    4 f   0.75707   0.50000   0.31543   1.00000 Dx,0,Dz
O3   O    4 d   0.25884   0.25884   0.39033   1.00000 Dx,Dx,Dz
O4   O    4 d   0.22854   0.22854   0.25008   1.00000 Dx,Dx,Dz
O5   O    4 f   0.80869   0.50000   0.17440   1.00000 Dx,0,Dz
O6   O    4 e   0.30125   0.00000   0.08799   1.00000 Dx,0,Dz
Xe1  Xe   2 c   0.50000   0.00000   0.12519   1.00000 0,0,Dz

# end of cif
```

# #(Tl₂)₃|(Na₄Mg₂F₈)|(OsO₄)

```
_cell_length_a      7.7953700000
_cell_length_b      7.7953700000
_cell_length_c      30.0000000000
_cell_angle_alpha   90.0000000000
_cell_angle_beta    90.0000000000
_cell_angle_gamma   90.0000000000
_cell_volume        1823.0338031070

_symmetry_space_group_name_H-M "P 4 m m"
_symmetry_Int_Tables_number 99
_space_group.reference_setting '099:P 4 -2'
_space_group.transform_Pp_abc a,b,c;0,0,0

loop_
_space_group_symop_id
_space_group_symop_operation_xyz
1 x,y,z
2 -x,-y,z
3 -y,x,z
4 y,-x,z
5 -x,y,z
6 x,-y,z
7 y,x,z
8 -y,-x,z

loop_
_atom_site_label
_atom_site_type_symbol
_atom_site_symmetry_multiplicity
_atom_site_Wyckoff_symbol
_atom_site_fract_x
_atom_site_fract_y
_atom_site_fract_z
_atom_site_occupancy
_atom_site_fract_symmform
Tl1  Tl  4 d  0.20222  0.20222  0.57181  1.00000 Dx,Dx,Dz
Tl2  Tl  2 c  0.50000  0.00000  0.51539  1.00000 0,0,Dz
Tl3  Tl  1 b  0.50000  0.50000  0.54096  1.00000 0,0,Dz
Tl4  Tl  1 a  0.00000  0.00000  0.47936  1.00000 0,0,Dz
Tl5  Tl  4 d  0.28649  0.28649  0.45754  1.00000 Dx,Dx,Dz
Na1  Na  2 c  0.50000  0.00000  0.36867  1.00000 0,0,Dz
Na2  Na  1 b  0.50000  0.50000  0.35907  1.00000 0,0,Dz
Na3  Na  1 a  0.00000  0.00000  0.36640  1.00000 0,0,Dz
Na4  Na  2 c  0.50000  0.00000  0.25060  1.00000 0,0,Dz
Na5  Na  1 b  0.50000  0.50000  0.23928  1.00000 0,0,Dz
Na6  Na  1 a  0.00000  0.00000  0.23539  1.00000 0,0,Dz
Mg1  Mg  4 d  0.24720  0.24720  0.30974  1.00000 Dx,Dx,Dz
F1   F   4 e  0.74380  0.00000  0.32693  1.00000 Dx,0,Dz
F2   F   4 f  0.75462  0.50000  0.29926  1.00000 Dx,0,Dz
F3   F   4 d  0.29437  0.29437  0.37437  1.00000 Dx,Dx,Dz
F4   F   4 d  0.20381  0.20381  0.24829  1.00000 Dx,Dx,Dz
O1   O   4 f  0.81962  0.50000  0.17655  1.00000 Dx,0,Dz
O2   O   4 e  0.31864  0.00000  0.10973  1.00000 Dx,0,Dz
Os1  Os  2 c  0.50000  0.00000  0.14251  1.00000 0,0,Dz

# end of cif
```

# #(Tl$_2$)$_3$|(Na$_4$Mg$_2$F$_8$)|(RuO$_4$)

```
_cell_length_a      7.7953700000
_cell_length_b      7.7953700000
_cell_length_c      30.0000000000
_cell_angle_alpha   90.0000000000
_cell_angle_beta    90.0000000000
_cell_angle_gamma   90.0000000000
_cell_volume        1823.0338031070

_symmetry_space_group_name_H-M "P 4 m m"
_symmetry_Int_Tables_number 99
_space_group.reference_setting '099:P 4 -2'
_space_group.transform_Pp_abc a,b,c;0,0,0

loop_
_space_group_symop_id
_space_group_symop_operation_xyz
1 x,y,z
2 -x,-y,z
3 -y,x,z
4 y,-x,z
5 -x,y,z
6 x,-y,z
7 y,x,z
8 -y,-x,z

loop_
_atom_site_label
_atom_site_type_symbol
_atom_site_symmetry_multiplicity
_atom_site_Wyckoff_symbol
_atom_site_fract_x
_atom_site_fract_y
_atom_site_fract_z
_atom_site_occupancy
_atom_site_fract_symmform
Tl1 Tl  4 d  0.20191  0.20191  0.57027  1.00000 Dx,Dx,Dz
Tl2 Tl  2 c  0.50000  0.00000  0.51365  1.00000 0,0,Dz
Tl3 Tl  1 b  0.50000  0.50000  0.53946  1.00000 0,0,Dz
Tl4 Tl  1 a  0.00000  0.00000  0.47801  1.00000 0,0,Dz
Tl5 Tl  4 d  0.28669  0.28669  0.45604  1.00000 Dx,Dx,Dz
Na1 Na  2 c  0.50000  0.00000  0.36668  1.00000 0,0,Dz
Na2 Na  1 b  0.50000  0.50000  0.35531  1.00000 0,0,Dz
Na3 Na  1 a  0.00000  0.00000  0.36531  1.00000 0,0,Dz
Na4 Na  2 c  0.50000  0.00000  0.25001  1.00000 0,0,Dz
Na5 Na  1 b  0.50000  0.50000  0.22943  1.00000 0,0,Dz
Na6 Na  1 a  0.00000  0.00000  0.23278  1.00000 0,0,Dz
Mg1 Mg  4 d  0.24679  0.24679  0.30936  1.00000 Dx,Dx,Dz
F1  F   4 e  0.74529  0.00000  0.32725  1.00000 Dx,0,Dz
F2  F   4 f  0.75742  0.50000  0.30058  1.00000 Dx,0,Dz
F3  F   4 d  0.29455  0.29455  0.37470  1.00000 Dx,Dx,Dz
F4  F   4 d  0.20330  0.20330  0.24831  1.00000 Dx,Dx,Dz
O1  O   4 f  0.82226  0.50000  0.17961  1.00000 Dx,0,Dz
O2  O   4 e  0.32117  0.00000  0.11366  1.00000 Dx,0,Dz
Ru1 Ru  2 c  0.50000  0.00000  0.14579  1.00000 0,0,Dz

# end of cif
```

**#(Tl₂)₃|(Na₄Mg₂F₈)|(XeO₄)**

```
_cell_length_a        7.7642200000
_cell_length_b        7.7642200000
_cell_length_c        30.0000000000
_cell_angle_alpha     90.0000000000
_cell_angle_beta      90.0000000000
_cell_angle_gamma     90.0000000000
_cell_volume          1808.4933662520

_symmetry_space_group_name_H-M "P 4 m m"
_symmetry_Int_Tables_number 99
_space_group.reference_setting '099:P 4 -2'
_space_group.transform_Pp_abc a,b,c;0,0,0

loop_
_space_group_symop_id
_space_group_symop_operation_xyz
1 x,y,z
2 -x,-y,z
3 -y,x,z
4 y,-x,z
5 -x,y,z
6 x,-y,z
7 y,x,z
8 -y,-x,z

loop_
_atom_site_label
_atom_site_type_symbol
_atom_site_symmetry_multiplicity
_atom_site_Wyckoff_symbol
_atom_site_fract_x
_atom_site_fract_y
_atom_site_fract_z
_atom_site_occupancy
_atom_site_fract_symmform
Tl1  Tl   4 d   0.20342   0.20342   0.57568   1.00000  Dx,Dx,Dz
Tl2  Tl   2 c   0.50000   0.00000   0.51688   1.00000  0,0,Dz
Tl3  Tl   1 b   0.50000   0.50000   0.54610   1.00000  0,0,Dz
Tl4  Tl   1 a   0.00000   0.00000   0.48589   1.00000  0,0,Dz
Tl5  Tl   4 d   0.28761   0.28761   0.45999   1.00000  Dx,Dx,Dz
Na1  Na   2 c   0.50000   0.00000   0.36570   1.00000  0,0,Dz
Na2  Na   1 b   0.50000   0.50000   0.35447   1.00000  0,0,Dz
Na3  Na   1 a   0.00000   0.00000   0.36980   1.00000  0,0,Dz
Na4  Na   2 c   0.50000   0.00000   0.24172   1.00000  0,0,Dz
Na5  Na   1 b   0.50000   0.50000   0.17842   1.00000  0,0,Dz
Na6  Na   1 a   0.00000   0.00000   0.23517   1.00000  0,0,Dz
Mg1  Mg   4 d   0.24524   0.24524   0.31267   1.00000  Dx,Dx,Dz
F1   F    4 e   0.75201   0.00000   0.33257   1.00000  Dx,0,Dz
F2   F    4 f   0.76484   0.50000   0.31120   1.00000  Dx,0,Dz
F3   F    4 d   0.29035   0.29035   0.38305   1.00000  Dx,Dx,Dz
F4   F    4 d   0.20287   0.20287   0.25211   1.00000  Dx,Dx,Dz
O1   O    4 f   0.81343   0.50000   0.17305   1.00000  Dx,0,Dz
O2   O    4 e   0.29079   0.00000   0.09770   1.00000  Dx,0,Dz
Xe1  Xe   2 c   0.50000   0.00000   0.13057   1.00000  0,0,Dz

# end of cif
```

# ☐|(Na₈Mg₄F₁₆)|(Cs₄Au₄Cl₁₂)

```
_cell_length_a      7.7953700000
_cell_length_b      7.7953700000
_cell_length_c      40.0000000000
_cell_angle_alpha   90.0000000000
_cell_angle_beta    90.0000000000
_cell_angle_gamma   90.0000000000
_cell_volume        2430.7117374760

_symmetry_space_group_name_H-M "P 4 m m"
_symmetry_Int_Tables_number 99
_space_group.reference_setting '099:P 4 -2'
_space_group.transform_Pp_abc a,b,c;0,0,0

loop_
_space_group_symop_id
_space_group_symop_operation_xyz
1 x,y,z
2 -x,-y,z
3 -y,x,z
4 y,-x,z
5 -x,y,z
6 x,-y,z
7 y,x,z
8 -y,-x,z

loop_
_atom_type_symbol
Cs
Au
Cl
Na
Mg
F

loop_
_atom_site_label
_atom_site_type_symbol
_atom_site_symmetry_multiplicity
_atom_site_Wyckoff_symbol
_atom_site_fract_x
_atom_site_fract_y
_atom_site_fract_z
_atom_site_occupancy
_atom_site_fract_symmform
Cs1 Cs   2 c   0.5000000000   0.0000000000   0.1193400000   1.0000000000  0,0,Dz
Cs2 Cs   2 c   0.5000000000   0.0000000000   0.2537900000   1.0000000000  0,0,Dz
Au1 Au   1 a   0.0000000000   0.0000000000   0.0549100000   1.0000000000  0,0,Dz
Au2 Au   1 b   0.5000000000   0.5000000000   0.1873500000   1.0000000000  0,0,Dz
Au3 Au   1 b   0.5000000000   0.5000000000   0.0553600000   1.0000000000  0,0,Dz
Au4 Au   1 a   0.0000000000   0.0000000000   0.3148600000   1.0000000000  0,0,Dz
Au5 Au   1 b   0.5000000000   0.5000000000   0.3246300000   1.0000000000  0,0,Dz
Au6 Au   1 a   0.0000000000   0.0000000000   0.1855800000   1.0000000000  0,0,Dz
Cl1 Cl   4 d   0.2107400000   0.2107400000   0.0565800000   1.0000000000  Dx,Dx,Dz
Cl2 Cl   4 d   0.2106800000   0.2106800000   0.3139400000   1.0000000000  Dx,Dx,Dz
Cl3 Cl   4 d   0.7107500000   0.7107500000   0.1872900000   1.0000000000  Dx,Dx,Dz
Cl4 Cl   1 b   0.5000000000   0.5000000000   0.1113700000   1.0000000000  0,0,Dz
Cl5 Cl   1 b   0.5000000000   0.5000000000   0.2689600000   1.0000000000  0,0,Dz
Cl6 Cl   1 a   0.0000000000   0.0000000000   0.2416700000   1.0000000000  0,0,Dz
Cl7 Cl   1 a   0.0000000000   0.0000000000   0.1294400000   1.0000000000  0,0,Dz
Na1 Na   4 d   0.2359000000   0.2359000000   0.3831700000   1.0000000000  Dx,Dx,Dz
Na2 Na   4 d   0.2505500000   0.2505500000   0.4718200000   1.0000000000  Dx,Dx,Dz
Mg1 Mg   1 b   0.5000000000   0.5000000000   0.4286000000   1.0000000000  0,0,Dz
Mg2 Mg   2 c   0.5000000000   0.0000000000   0.4279900000   1.0000000000  0,0,Dz
```

```
Mg3  Mg    1 a  0.0000000000  0.0000000000  0.4308700000  1.0000000000 0,0,Dz
F1   F     4 f  0.2521900000  0.5000000000  0.4271500000  1.0000000000 Dx,0,Dz
F2   F     4 e  0.2493300000  0.0000000000  0.4294600000  1.0000000000 Dx,0,Dz
F3   F     1 b  0.5000000000  0.5000000000  0.3758700000  1.0000000000 0,0,Dz
F4   F     2 c  0.5000000000  0.0000000000  0.3797500000  1.0000000000 0,0,Dz
F5   F     1 a  0.0000000000  0.0000000000  0.3816600000  1.0000000000 0,0,Dz
F6   F     1 b  0.5000000000  0.5000000000  0.4773800000  1.0000000000 0,0,Dz
F7   F     2 c  0.5000000000  0.0000000000  0.4771000000  1.0000000000 0,0,Dz
F8   F     1 a  0.0000000000  0.0000000000  0.4793300000  1.0000000000 0,0,Dz

# end of cif
```

# #(Li₄)|(Na₈Mg₄F₁₆)|(Cs₄Au₄Cl₁₂)

```
_cell_length_a       7.7953700000
_cell_length_b       7.7953700000
_cell_length_c       40.0000000000
_cell_angle_alpha    90.0000000000
_cell_angle_beta     90.0000000000
_cell_angle_gamma    90.0000000000
_cell_volume         2430.7117374760

_symmetry_space_group_name_H-M "P 4 m m"
_symmetry_Int_Tables_number 99
_space_group.reference_setting '099:P 4 -2'
_space_group.transform_Pp_abc a,b,c;0,0,0

loop_
_space_group_symop_id
_space_group_symop_operation_xyz
1 x,y,z
2 -x,-y,z
3 -y,x,z
4 y,-x,z
5 -x,y,z
6 x,-y,z
7 y,x,z
8 -y,-x,z

loop_
_atom_type_symbol
Cs
Au
Cl
Na
Mg
F
Li

loop_
_atom_site_label
_atom_site_type_symbol
_atom_site_symmetry_multiplicity
_atom_site_Wyckoff_symbol
_atom_site_fract_x
_atom_site_fract_y
_atom_site_fract_z
_atom_site_occupancy
_atom_site_fract_symmform
Cs1 Cs   2 c   0.5000000000   0.0000000000   0.1173700000   1.0000000000 0,0,Dz
Cs2 Cs   2 c   0.5000000000   0.0000000000   0.2512700000   1.0000000000 0,0,Dz
Au1 Au   1 a   0.0000000000   0.0000000000   0.0538900000   1.0000000000 0,0,Dz
Au2 Au   1 b   0.5000000000   0.5000000000   0.1872900000   1.0000000000 0,0,Dz
Au3 Au   1 b   0.5000000000   0.5000000000   0.0557600000   1.0000000000 0,0,Dz
Au4 Au   1 a   0.0000000000   0.0000000000   0.3119000000   1.0000000000 0,0,Dz
Au5 Au   1 b   0.5000000000   0.5000000000   0.3245000000   1.0000000000 0,0,Dz
Au6 Au   1 a   0.0000000000   0.0000000000   0.1847500000   1.0000000000 0,0,Dz
Cl1 Cl   4 d   0.2114900000   0.2114900000   0.0554000000   1.0000000000 Dx,Dx,Dz
Cl2 Cl   4 d   0.2190100000   0.2190100000   0.3121200000   1.0000000000 Dx,Dx,Dz
Cl3 Cl   4 d   0.7109100000   0.7109100000   0.1872000000   1.0000000000 Dx,Dx,Dz
Cl4 Cl   1 b   0.5000000000   0.5000000000   0.1121100000   1.0000000000 0,0,Dz
Cl5 Cl   1 b   0.5000000000   0.5000000000   0.2684500000   1.0000000000 0,0,Dz
Cl6 Cl   1 a   0.0000000000   0.0000000000   0.2409900000   1.0000000000 0,0,Dz
Cl7 Cl   1 a   0.0000000000   0.0000000000   0.1286500000   1.0000000000 0,0,Dz
Na1 Na   4 d   0.2329000000   0.2329000000   0.3765800000   1.0000000000 Dx,Dx,Dz
Na2 Na   4 d   0.2538300000   0.2538300000   0.4693300000   1.0000000000 Dx,Dx,Dz
Mg1 Mg   1 b   0.5000000000   0.5000000000   0.4281100000   1.0000000000 0,0,Dz
```

```
Mg2  Mg    2 c   0.5000000000   0.0000000000   0.4301600000   1.0000000000  0,0,Dz
Mg3  Mg    1 a   0.0000000000   0.0000000000   0.4332000000   1.0000000000  0,0,Dz
F1   F     4 f   0.2518900000   0.5000000000   0.4326700000   1.0000000000  Dx,0,Dz
F2   F     4 e   0.2502400000   0.0000000000   0.4352100000   1.0000000000  Dx,0,Dz
F3   F     1 b   0.5000000000   0.5000000000   0.3785400000   1.0000000000  0,0,Dz
F4   F     2 c   0.5000000000   0.0000000000   0.3830000000   1.0000000000  0,0,Dz
F5   F     1 a   0.0000000000   0.0000000000   0.3854700000   1.0000000000  0,0,Dz
F6   F     1 b   0.5000000000   0.5000000000   0.4809700000   1.0000000000  0,0,Dz
F7   F     2 c   0.5000000000   0.0000000000   0.4819400000   1.0000000000  0,0,Dz
F8   F     1 a   0.0000000000   0.0000000000   0.4837000000   1.0000000000  0,0,Dz
Li1  Li    2 c   0.5000000000   0.0000000000   0.5250100000   1.0000000000  0,0,Dz
Li2  Li    1 b   0.5000000000   0.5000000000   0.5242000000   1.0000000000  0,0,Dz
Li3  Li    1 a   0.0000000000   0.0000000000   0.5266100000   1.0000000000  0,0,Dz

# end of cif
```

# #(Li$_4$)$_3$|(Na$_8$Mg$_4$F$_{16}$)|(Cs$_4$Au$_4$Cl$_{12}$)

```
_cell_length_a      7.7953700000
_cell_length_b      7.7953700000
_cell_length_c      50.0000000000
_cell_angle_alpha   90.0000000000
_cell_angle_beta    90.0000000000
_cell_angle_gamma   90.0000000000
_cell_volume        3038.3896718450

_symmetry_space_group_name_H-M "P 4 m m"
_symmetry_Int_Tables_number 99
_space_group.reference_setting '099:P 4 -2'
_space_group.transform_Pp_abc a,b,c;0,0,0

loop_
_space_group_symop_id
_space_group_symop_operation_xyz
1 x,y,z
2 -x,-y,z
3 -y,x,z
4 y,-x,z
5 -x,y,z
6 x,-y,z
7 y,x,z
8 -y,-x,z

loop_
_atom_type_symbol
Cs
Au
Cl
Na
Mg
F
Li

loop_
_atom_site_label
_atom_site_type_symbol
_atom_site_symmetry_multiplicity
_atom_site_Wyckoff_symbol
_atom_site_fract_x
_atom_site_fract_y
_atom_site_fract_z
_atom_site_occupancy
_atom_site_fract_symmform
Cs1 Cs   2 c   0.5000000000   0.0000000000   0.0952700000   1.0000000000 0,0,Dz
Cs2 Cs   2 c   0.5000000000   0.0000000000   0.2025000000   1.0000000000 0,0,Dz
Au1 Au   1 a   0.0000000000   0.0000000000   0.0444500000   1.0000000000 0,0,Dz
Au2 Au   1 b   0.5000000000   0.5000000000   0.1510800000   1.0000000000 0,0,Dz
Au3 Au   1 b   0.5000000000   0.5000000000   0.0456400000   1.0000000000 0,0,Dz
Au4 Au   1 a   0.0000000000   0.0000000000   0.2510400000   1.0000000000 0,0,Dz
Au5 Au   1 b   0.5000000000   0.5000000000   0.2610900000   1.0000000000 0,0,Dz
Au6 Au   1 a   0.0000000000   0.0000000000   0.1491500000   1.0000000000 0,0,Dz
Cl1 Cl   4 d   0.2113800000   0.2113800000   0.0456800000   1.0000000000 Dx,Dx,Dz
Cl2 Cl   4 d   0.2182500000   0.2182500000   0.2512100000   1.0000000000 Dx,Dx,Dz
Cl3 Cl   4 d   0.7108800000   0.7108800000   0.1510400000   1.0000000000 Dx,Dx,Dz
Cl4 Cl   1 b   0.5000000000   0.5000000000   0.0907000000   1.0000000000 0,0,Dz
Cl5 Cl   1 b   0.5000000000   0.5000000000   0.2163200000   1.0000000000 0,0,Dz
Cl6 Cl   1 a   0.0000000000   0.0000000000   0.1941500000   1.0000000000 0,0,Dz
Cl7 Cl   1 a   0.0000000000   0.0000000000   0.1042900000   1.0000000000 0,0,Dz
Na1 Na   4 d   0.2328100000   0.2328100000   0.3030600000   1.0000000000 Dx,Dx,Dz
Na2 Na   4 d   0.2535800000   0.2535800000   0.3770100000   1.0000000000 Dx,Dx,Dz
Mg1 Mg   1 b   0.5000000000   0.5000000000   0.3438400000   1.0000000000 0,0,Dz
```

```
Mg2  Mg   2 c   0.5000000000   0.0000000000   0.3452800000   1.0000000000  0,0,Dz
Mg3  Mg   1 a   0.0000000000   0.0000000000   0.3476600000   1.0000000000  0,0,Dz
F1   F    4 f   0.2518400000   0.5000000000   0.3471200000   1.0000000000  Dx,0,Dz
F2   F    4 e   0.2501500000   0.0000000000   0.3490600000   1.0000000000  Dx,0,Dz
F3   F    1 b   0.5000000000   0.5000000000   0.3040500000   1.0000000000  0,0,Dz
F4   F    2 c   0.5000000000   0.0000000000   0.3074900000   1.0000000000  0,0,Dz
F5   F    1 a   0.0000000000   0.0000000000   0.3094100000   1.0000000000  0,0,Dz
F6   F    1 b   0.5000000000   0.5000000000   0.3855600000   1.0000000000  0,0,Dz
F7   F    2 c   0.5000000000   0.0000000000   0.3862300000   1.0000000000  0,0,Dz
F8   F    1 a   0.0000000000   0.0000000000   0.3876500000   1.0000000000  0,0,Dz
Li1  Li   2 c   0.5000000000   0.0000000000   0.4209700000   1.0000000000  0,0,Dz
Li2  Li   1 b   0.5000000000   0.5000000000   0.4204400000   1.0000000000  0,0,Dz
Li3  Li   1 a   0.0000000000   0.0000000000   0.4222200000   1.0000000000  0,0,Dz
Li4  Li   4 d   0.2515400000   0.2515400000   0.4546800000   1.0000000000  Dx,Dx,Dz
Li5  Li   2 c   0.5000000000   0.0000000000   0.4827900000   1.0000000000  0,0,Dz
Li6  Li   1 b   0.5000000000   0.5000000000   0.4829000000   1.0000000000  0,0,Dz
Li7  Li   1 a   0.0000000000   0.0000000000   0.4832200000   1.0000000000  0,0,Dz

# end of cif
```

# **#(Cs₄Au₄Cl₁₂)|(Na₈Mg₄F₁₆)|(Xe₄F₈)**

```
_cell_length_a       7.7953700000
_cell_length_b       7.7953700000
_cell_length_c       50.0000000000
_cell_angle_alpha    90.0000000000
_cell_angle_beta     90.0000000000
_cell_angle_gamma    90.0000000000
_cell_volume         3038.3896718450

_symmetry_space_group_name_H-M "P 4 m m"
_symmetry_Int_Tables_number 99
_space_group.reference_setting '099:P 4 -2'
_space_group.transform_Pp_abc a,b,c;0,0,0

loop_
_space_group_symop_id
_space_group_symop_operation_xyz
1 x,y,z
2 -x,-y,z
3 -y,x,z
4 y,-x,z
5 -x,y,z
6 x,-y,z
7 y,x,z
8 -y,-x,z

loop_
_atom_type_symbol
Cs
Au
Cl
Na
Mg
F
Xe

loop_
_atom_site_label
_atom_site_type_symbol
_atom_site_symmetry_multiplicity
_atom_site_Wyckoff_symbol
_atom_site_fract_x
_atom_site_fract_y
_atom_site_fract_z
_atom_site_occupancy
_atom_site_fract_symmform
Cs1  Cs   2  c   0.5000000000   0.0000000000   0.0972800000   1.0000000000  0,0,Dz
Cs2  Cs   2  c   0.5000000000   0.0000000000   0.2050000000   1.0000000000  0,0,Dz
Au1  Au   1  a   0.0000000000   0.0000000000   0.0456600000   1.0000000000  0,0,Dz
Au2  Au   1  b   0.5000000000   0.5000000000   0.1515900000   1.0000000000  0,0,Dz
Au3  Au   1  b   0.5000000000   0.5000000000   0.0460700000   1.0000000000  0,0,Dz
Au4  Au   1  a   0.0000000000   0.0000000000   0.2534400000   1.0000000000  0,0,Dz
Au5  Au   1  b   0.5000000000   0.5000000000   0.2602600000   1.0000000000  0,0,Dz
Au6  Au   1  a   0.0000000000   0.0000000000   0.1500700000   1.0000000000  0,0,Dz
Cl1  Cl   4  d   0.2107400000   0.2107400000   0.0470500000   1.0000000000  Dx,Dx,Dz
Cl2  Cl   4  d   0.2107600000   0.2107600000   0.2525900000   1.0000000000  Dx,Dx,Dz
Cl3  Cl   4  d   0.7107200000   0.7107200000   0.1515600000   1.0000000000  Dx,Dx,Dz
Cl4  Cl   1  b   0.5000000000   0.5000000000   0.0908800000   1.0000000000  0,0,Dz
Cl5  Cl   1  b   0.5000000000   0.5000000000   0.2157600000   1.0000000000  0,0,Dz
Cl6  Cl   1  a   0.0000000000   0.0000000000   0.1948800000   1.0000000000  0,0,Dz
Cl7  Cl   1  a   0.0000000000   0.0000000000   0.1052100000   1.0000000000  0,0,Dz
Na1  Na   4  d   0.2338500000   0.2338500000   0.3080200000   1.0000000000  Dx,Dx,Dz
Na2  Na   4  d   0.2563200000   0.2563200000   0.3780200000   1.0000000000  Dx,Dx,Dz
Mg1  Mg   1  b   0.5000000000   0.5000000000   0.3425600000   1.0000000000  0,0,Dz
```

```
Mg2  Mg    2 c   0.5000000000   0.0000000000   0.3434400000   1.0000000000   0,0,Dz
Mg3  Mg    1 a   0.0000000000   0.0000000000   0.3468200000   1.0000000000   0,0,Dz
F1   F     4 f   0.2519800000   0.5000000000   0.3419700000   1.0000000000   Dx,0,Dz
F2   F     4 e   0.2491050000   0.0000000000   0.3448400000   1.0000000000   Dx,0,Dz
F3   F     1 b   0.5000000000   0.5000000000   0.3011100000   1.0000000000   0,0,Dz
F4   F     2 c   0.5000000000   0.0000000000   0.3047500000   1.0000000000   0,0,Dz
F5   F     1 a   0.0000000000   0.0000000000   0.3069400000   1.0000000000   0,0,Dz
F6   F     1 b   0.5000000000   0.5000000000   0.3813000000   1.0000000000   0,0,Dz
F7   F     2 c   0.5000000000   0.0000000000   0.3826600000   1.0000000000   0,0,Dz
F8   F     1 a   0.0000000000   0.0000000000   0.3858100000   1.0000000000   0,0,Dz
F9   F     4 d   0.3077100000   0.3077100000   0.4223500000   1.0000000000   Dx,Dx,Dz
F10  F     4 d   0.1977000000   0.1977000000   0.5002900000   1.0000000000   Dx,Dx,Dz
Xe1  Xe    4 d   0.2497700000   0.2497700000   0.4612300000   1.0000000000   Dx,Dx,Dz

# end of cif
```

# ▢|(Li₂F₂)₃|(CaCuO₂)

```
_cell_length_a       3.9533491608
_cell_length_b       3.9533491608
_cell_length_c       30.0000000000
_cell_angle_alpha    90.0000000000
_cell_angle_beta     90.0000000000
_cell_angle_gamma    90.0000000000
_cell_volume         468.8690876160

_symmetry_space_group_name_H-M "P 4 m m"
_symmetry_Int_Tables_number 99
_space_group.reference_setting '099:P 4 -2'
_space_group.transform_Pp_abc a,b,c;0,0,0

loop_
_space_group_symop_id
_space_group_symop_operation_xyz
1 x,y,z
2 -x,-y,z
3 -y,x,z
4 y,-x,z
5 -x,y,z
6 x,-y,z
7 y,x,z
8 -y,-x,z

loop_
_atom_type_symbol
Cu
O
Ca
Li
F

loop_
_atom_site_label
_atom_site_type_symbol
_atom_site_symmetry_multiplicity
_atom_site_Wyckoff_symbol
_atom_site_fract_x
_atom_site_fract_y
_atom_site_fract_z
_atom_site_occupancy
_atom_site_fract_symmform
Cu1  Cu   1 a   0.0000000000   0.0000000000   0.8318900000   1.0000000000   0,0,Dz
O1   O    2 c   0.5000000000   0.0000000000   0.8374500000   1.0000000000   0,0,Dz
Ca1  Ca   1 b   0.5000000000   0.5000000000   0.8640500000   1.0000000000   0,0,Dz
Li1  Li   2 c   0.5000000000   0.0000000000   0.6295900000   1.0000000000   0,0,Dz
Li2  Li   1 b   0.5000000000   0.5000000000   0.6951200000   1.0000000000   0,0,Dz
Li3  Li   1 a   0.0000000000   0.0000000000   0.6936800000   1.0000000000   0,0,Dz
Li4  Li   2 c   0.5000000000   0.0000000000   0.7614300000   1.0000000000   0,0,Dz
F1   F    1 b   0.5000000000   0.5000000000   0.6268100000   1.0000000000   0,0,Dz
F2   F    1 a   0.0000000000   0.0000000000   0.6268000000   1.0000000000   0,0,Dz
F3   F    2 c   0.5000000000   0.0000000000   0.6944000000   1.0000000000   0,0,Dz
F4   F    1 b   0.5000000000   0.5000000000   0.7602200000   1.0000000000   0,0,Dz
F5   F    1 a   0.0000000000   0.0000000000   0.7600200000   1.0000000000   0,0,Dz

# end of cif
```

# **(Li$_2$)|(Li$_2$F$_2$)$_3$|(CaCuO$_2$)**

```
_cell_length_a       3.9533498679
_cell_length_b       3.9533498679
_cell_length_c       30.0000000000
_cell_angle_alpha    90.0000000000
_cell_angle_beta     90.0000000000
_cell_angle_gamma    90.0000000000
_cell_volume         468.8692553424

_symmetry_space_group_name_H-M "P 4 m m"
_symmetry_Int_Tables_number 99
_space_group.reference_setting '099:P 4 -2'
_space_group.transform_Pp_abc a,b,c;0,0,0

loop_
_space_group_symop_id
_space_group_symop_operation_xyz
1 x,y,z
2 -x,-y,z
3 -y,x,z
4 y,-x,z
5 -x,y,z
6 x,-y,z
7 y,x,z
8 -y,-x,z

loop_
_atom_type_symbol
Cu
O
Ca
Li
F

loop_
_atom_site_label
_atom_site_type_symbol
_atom_site_symmetry_multiplicity
_atom_site_Wyckoff_symbol
_atom_site_fract_x
_atom_site_fract_y
_atom_site_fract_z
_atom_site_occupancy
_atom_site_fract_symmform
Cu1  Cu   1 a   0.0000000000   0.0000000000   0.8425320000   1.0000000000  0,0,Dz
O1   O    2 c   0.5000000000   0.0000000000   0.8423480000   1.0000000000  0,0,Dz
Ca1  Ca   1 b   0.5000000000   0.5000000000   0.8720560000   1.0000000000  0,0,Dz
Li1  Li   2 c   0.5000000000   0.0000000000   0.6256900000   1.0000000000  0,0,Dz
Li2  Li   1 b   0.5000000000   0.5000000000   0.7008840000   1.0000000000  0,0,Dz
Li3  Li   1 a   0.0000000000   0.0000000000   0.7000520000   1.0000000000  0,0,Dz
Li4  Li   2 c   0.5000000000   0.0000000000   0.7767910000   1.0000000000  0,0,Dz
Li5  Li   1 b   0.5000000000   0.5000000000   0.5491610000   1.0000000000  0,0,Dz
Li6  Li   1 a   0.0000000000   0.0000000000   0.5501830000   1.0000000000  0,0,Dz
F1   F    1 b   0.5000000000   0.5000000000   0.6104670000   1.0000000000  0,0,Dz
F2   F    1 a   0.0000000000   0.0000000000   0.6112510000   1.0000000000  0,0,Dz
F3   F    2 c   0.5000000000   0.0000000000   0.6862600000   1.0000000000  0,0,Dz
F4   F    1 b   0.5000000000   0.5000000000   0.7612870000   1.0000000000  0,0,Dz
F5   F    1 a   0.0000000000   0.0000000000   0.7607290000   1.0000000000  0,0,Dz

# end of cif
```

# #(CaCuO$_2$)|(Li$_2$F$_2$)$_3$|(XeF$_2$)

```
_cell_length_a        3.9533498679
_cell_length_b        3.9533498679
_cell_length_c        40.0000000000
_cell_angle_alpha     90.0000000000
_cell_angle_beta      90.0000000000
_cell_angle_gamma     90.0000000000
_cell_volume          625.1590071232

_symmetry_space_group_name_H-M "P m m 2"
_symmetry_Int_Tables_number 25
_space_group.reference_setting '025:P 2 -2'
_space_group.transform_Pp_abc a,b,c;0,0,0

loop_
_space_group_symop_id
_space_group_symop_operation_xyz
1 x,y,z
2 -x,-y,z
3 -x,y,z
4 x,-y,z

loop_
_atom_type_symbol
Cu
O
Ca
Li
F
Xe

loop_
_atom_site_label
_atom_site_type_symbol
_atom_site_symmetry_multiplicity
_atom_site_Wyckoff_symbol
_atom_site_fract_x
_atom_site_fract_y
_atom_site_fract_z
_atom_site_occupancy
_atom_site_fract_symmform
Cu1  Cu   1 a   0.0000000000   0.0000000000   0.6247670000   1.0000000000  0,0,Dz
O1   O    1 b   0.0000000000   0.5000000000   0.6293140000   1.0000000000  0,0,Dz
O2   O    1 c   0.5000000000   0.0000000000   0.6293900000   1.0000000000  0,0,Dz
Ca1  Ca   1 d   0.5000000000   0.5000000000   0.6491540000   1.0000000000  0,0,Dz
Li1  Li   1 b   0.0000000000   0.5000000000   0.4737230000   1.0000000000  0,0,Dz
Li2  Li   1 c   0.5000000000   0.0000000000   0.4722310000   1.0000000000  0,0,Dz
Li3  Li   1 d   0.5000000000   0.5000000000   0.5222320000   1.0000000000  0,0,Dz
Li4  Li   1 a   0.0000000000   0.0000000000   0.5211220000   1.0000000000  0,0,Dz
Li5  Li   1 b   0.0000000000   0.5000000000   0.5718070000   1.0000000000  0,0,Dz
Li6  Li   1 c   0.5000000000   0.0000000000   0.5715000000   1.0000000000  0,0,Dz
F1   F    1 d   0.5000000000   0.5000000000   0.4720450000   1.0000000000  0,0,Dz
F2   F    1 a   0.0000000000   0.0000000000   0.4719160000   1.0000000000  0,0,Dz
F3   F    1 b   0.0000000000   0.5000000000   0.5223450000   1.0000000000  0,0,Dz
F4   F    1 c   0.5000000000   0.0000000000   0.5221290000   1.0000000000  0,0,Dz
F5   F    1 d   0.5000000000   0.5000000000   0.5715920000   1.0000000000  0,0,Dz
F6   F    1 a   0.0000000000   0.0000000000   0.5713530000   1.0000000000  0,0,Dz
F7   F    1 c   0.5000000000   0.0000000000   0.4157290000   1.0000000000  0,0,Dz
F8   F    1 c   0.5000000000   0.0000000000   0.3133630000   1.0000000000  0,0,Dz
Xe1  Xe   1 c   0.5000000000   0.0000000000   0.3645340000   1.0000000000  0,0,Dz

# end of cif
```

# #(CaCuO$_2$)|(Li$_2$F$_2$)$_3$|(KrF$_2$)

```
_cell_length_a      3.9533498679
_cell_length_b      3.9533498679
_cell_length_c      40.0000000000
_cell_angle_alpha   90.0000000000
_cell_angle_beta    90.0000000000
_cell_angle_gamma   90.0000000000
_cell_volume        625.1590071232

_symmetry_space_group_name_H-M "P m m 2"
_symmetry_Int_Tables_number 25
_space_group.reference_setting '025:P 2 -2'
_space_group.transform_Pp_abc a,b,c;0,0,0

loop_
_space_group_symop_id
_space_group_symop_operation_xyz
1 x,y,z
2 -x,-y,z
3 -x,y,z
4 x,-y,z

loop_
_atom_type_symbol
Cu
O
Ca
Li
F
Kr

loop_
_atom_site_label
_atom_site_type_symbol
_atom_site_symmetry_multiplicity
_atom_site_Wyckoff_symbol
_atom_site_fract_x
_atom_site_fract_y
_atom_site_fract_z
_atom_site_occupancy
_atom_site_fract_symmform
Cu1  Cu   1 a   0.0000000000   0.0000000000   0.6245280000   1.0000000000  0,0,Dz
O1   O    1 b   0.0000000000   0.5000000000   0.6292430000   1.0000000000  0,0,Dz
O2   O    1 c   0.5000000000   0.0000000000   0.6292960000   1.0000000000  0,0,Dz
Ca1  Ca   1 d   0.5000000000   0.5000000000   0.6490540000   1.0000000000  0,0,Dz
Li1  Li   1 b   0.0000000000   0.5000000000   0.4729570000   1.0000000000  0,0,Dz
Li2  Li   1 c   0.5000000000   0.0000000000   0.4719200000   1.0000000000  0,0,Dz
Li3  Li   1 d   0.5000000000   0.5000000000   0.5217650000   1.0000000000  0,0,Dz
Li4  Li   1 a   0.0000000000   0.0000000000   0.5206400000   1.0000000000  0,0,Dz
Li5  Li   1 b   0.0000000000   0.5000000000   0.5712530000   1.0000000000  0,0,Dz
Li6  Li   1 c   0.5000000000   0.0000000000   0.5710940000   1.0000000000  0,0,Dz
F1   F    1 d   0.5000000000   0.5000000000   0.4717670000   1.0000000000  0,0,Dz
F2   F    1 a   0.0000000000   0.0000000000   0.4716140000   1.0000000000  0,0,Dz
F3   F    1 b   0.0000000000   0.5000000000   0.5219960000   1.0000000000  0,0,Dz
F4   F    1 c   0.5000000000   0.0000000000   0.5218800000   1.0000000000  0,0,Dz
F5   F    1 d   0.5000000000   0.5000000000   0.5713730000   1.0000000000  0,0,Dz
F6   F    1 a   0.0000000000   0.0000000000   0.5711570000   1.0000000000  0,0,Dz
F7   F    1 c   0.5000000000   0.0000000000   0.4143090000   1.0000000000  0,0,Dz
F8   F    1 c   0.5000000000   0.0000000000   0.3182110000   1.0000000000  0,0,Dz
Kr1  Kr   1 c   0.5000000000   0.0000000000   0.3661900000   1.0000000000  0,0,Dz

# end of cif
```

# #(CaCuO$_2$)|(Li$_2$F$_2$)$_3$|(F$_2$)

```
_cell_length_a       3.9533498679
_cell_length_b       3.9533498679
_cell_length_c       30.0000000000
_cell_angle_alpha    90.0000000000
_cell_angle_beta     90.0000000000
_cell_angle_gamma    90.0000000000
_cell_volume         468.8692553424

_symmetry_space_group_name_H-M "P 4 m m"
_symmetry_Int_Tables_number 99
_space_group.reference_setting '099:P 4 -2'
_space_group.transform_Pp_abc a,b,c;0,0,0

loop_
_space_group_symop_id
_space_group_symop_operation_xyz
1 x,y,z
2 -x,-y,z
3 -y,x,z
4 y,-x,z
5 -x,y,z
6 x,-y,z
7 y,x,z
8 -y,-x,z

loop_
_atom_type_symbol
Cu
O
Ca
Li
F

loop_
_atom_site_label
_atom_site_type_symbol
_atom_site_symmetry_multiplicity
_atom_site_Wyckoff_symbol
_atom_site_fract_x
_atom_site_fract_y
_atom_site_fract_z
_atom_site_occupancy
_atom_site_fract_symmform
Cu1  Cu   1 a   0.0000000000   0.0000000000   0.8376080000   1.0000000000  0,0,Dz
O1   O    2 c   0.5000000000   0.0000000000   0.8494380000   1.0000000000  0,0,Dz
Ca1  Ca   1 b   0.5000000000   0.5000000000   0.8753210000   1.0000000000  0,0,Dz
Li1  Li   2 c   0.5000000000   0.0000000000   0.6134160000   1.0000000000  0,0,Dz
Li2  Li   1 b   0.5000000000   0.5000000000   0.6907770000   1.0000000000  0,0,Dz
Li3  Li   1 a   0.0000000000   0.0000000000   0.6887340000   1.0000000000  0,0,Dz
Li4  Li   2 c   0.5000000000   0.0000000000   0.7606890000   1.0000000000  0,0,Dz
F1   F    1 b   0.5000000000   0.5000000000   0.6277610000   1.0000000000  0,0,Dz
F2   F    1 a   0.0000000000   0.0000000000   0.6268170000   1.0000000000  0,0,Dz
F3   F    2 c   0.5000000000   0.0000000000   0.6988050000   1.0000000000  0,0,Dz
F4   F    1 b   0.5000000000   0.5000000000   0.7690960000   1.0000000000  0,0,Dz
F5   F    1 a   0.0000000000   0.0000000000   0.7696580000   1.0000000000  0,0,Dz
F6   F    2 c   0.5000000000   0.0000000000   0.5511560000   1.0000000000  0,0,Dz

# end of cif
```

#(KMgF₃)|(NaCl)|(KMgF₃)

```
_cell_length_a                      3.973940
_cell_length_b                      3.973940
_cell_length_c                      25.620001
_cell_angle_alpha                   90.000000
_cell_angle_beta                    90.000000
_cell_angle_gamma                   90.000000
_cell_volume                        404.596182
_space_group_name_H-M_alt           'P 1'
_space_group_IT_number              1

loop_
_space_group_symop_operation_xyz
   'x, y, z'

loop_
   _atom_site_label
   _atom_site_occupancy
   _atom_site_fract_x
   _atom_site_fract_y
   _atom_site_fract_z
   _atom_site_adp_type
   _atom_site_U_iso_or_equiv
   _atom_site_type_symbol
   K1        1.0      0.000000     0.000000     0.390320     Uiso   ? K
   K2        1.0      0.000000     0.000000     0.609680     Uiso   ? K
   Na1       1.0      0.500000     0.500000     0.500000     Uiso   ? Na
   Cl1       1.0      0.000000     0.000000     0.500000     Uiso   ? Cl
   F1        1.0      0.500000     0.500000     0.390320     Uiso   ? F
   F2        1.0      0.500000     0.500000     0.609680     Uiso   ? F
   F3        1.0      0.000000     0.500000     0.312531     Uiso   ? F
   F4        1.0      0.000000     0.500000     0.687469     Uiso   ? F
   F5        1.0      0.500000     0.000000     0.312531     Uiso   ? F
   F6        1.0      0.500000     0.000000     0.687469     Uiso   ? F
   Mg1       1.0      0.500000     0.500000     0.687469     Uiso   ? Mg
   Mg2       1.0      0.500000     0.500000     0.312531     Uiso   ? Mg
```

# ☐ | (C₂) | ☐

```
_cell_length_a       2.4677240000
_cell_length_b       2.4677240000
_cell_length_c       14.3425190000
_cell_angle_alpha    90.0000000000
_cell_angle_beta     90.0000000000
_cell_angle_gamma    120.0000000000
_cell_volume         75.6396020518

_symmetry_space_group_name_H-M "P 6/m 2/m 2/m"
_symmetry_Int_Tables_number 191
_space_group.reference_setting '191:-P 6 2'
_space_group.transform_Pp_abc a,b,c;0,0,0

loop_
_space_group_symop_id
_space_group_symop_operation_xyz
1 x,y,z
2 x-y,x,z
3 -y,x-y,z
4 -x,-y,z
5 -x+y,-x,z
6 y,-x+y,z
7 x-y,-y,-z
8 x,x-y,-z
9 y,x,-z
10 -x+y,y,-z
11 -x,-x+y,-z
12 -y,-x,-z
13 -x,-y,-z
14 -x+y,-x,-z
15 y,-x+y,-z
16 x,y,-z
17 x-y,x,-z
18 -y,x-y,-z
19 -x+y,y,z
20 -x,-x+y,z
21 -y,-x,z
22 x-y,-y,z
23 x,x-y,z
24 y,x,z

loop_
_atom_site_label
_atom_site_type_symbol
_atom_site_symmetry_multiplicity
_atom_site_Wyckoff_symbol
_atom_site_fract_x
_atom_site_fract_y
_atom_site_fract_z
_atom_site_occupancy
_atom_site_fract_symmform
C1  C   2  d   0.33333   0.66667   0.50000   1.00000 0,0,0

# end of cif
```

# ▢|(Mg₃Cu₇H₄)|▢

```
_cell_length_a         4.0100000000
_cell_length_b         4.0100000000
_cell_length_c         32.0800000000
_cell_angle_alpha      90.0000000000
_cell_angle_beta       90.0000000000
_cell_angle_gamma      90.0000000000
_cell_volume           515.8496080000

_symmetry_space_group_name_H-M "P 4/m 2/m 2/m"
_symmetry_Int_Tables_number 123
_space_group.reference_setting '123:-P 4 2'
_space_group.transform_Pp_abc a,b,c;0,0,0

loop_
_space_group_symop_id
_space_group_symop_operation_xyz
1 x,y,z
2 x,-y,-z
3 -x,y,-z
4 -x,-y,z
5 -y,-x,-z
6 -y,x,z
7 y,-x,z
8 y,x,-z
9 -x,-y,-z
10 -x,y,z
11 x,-y,z
12 x,y,-z
13 y,x,z
14 y,-x,-z
15 -y,x,-z
16 -y,-x,z

loop_
_atom_type_symbol
Mg
Cu
H

loop_
_atom_site_label
_atom_site_type_symbol
_atom_site_symmetry_multiplicity
_atom_site_Wyckoff_symbol
_atom_site_fract_x
_atom_site_fract_y
_atom_site_fract_z
_atom_site_occupancy
_atom_site_fract_symmform
Mg1  Mg   2 g   0.0000000000   0.0000000000   0.3750000000   1.0000000000   0,0,Dz
Mg2  Mg   1 b   0.0000000000   0.0000000000   0.5000000000   1.0000000000   0,0,0
Cu1  Cu   2 h   0.5000000000   0.5000000000   0.3750000000   1.0000000000   0,0,Dz
Cu2  Cu   4 i   0.0000000000   0.5000000000   0.4375000000   1.0000000000   0,0,Dz
Cu3  Cu   1 d   0.5000000000   0.5000000000   0.5000000000   1.0000000000   0,0,0
H1   H    2 h   0.5000000000   0.5000000000   0.4375000000   1.0000000000   0,0,Dz

# end of cif
```

### 4. Superconductivity parameters for NaCl layers (@ degauss 0.02).

| System: (KMgF$_3$) | (NaCl) | (KMgF$_3$) | λ | N(Ef) [states/Ry] | T$_C$ [K] |
|---|---|---|---|
| 20% electron doping | 0.3510 | 19.1898 | 0.216 |
| 10% hole doping | 0.3138 | 114.8306 | 0.068 |

### 5. Superconductivity parameters for Mg$_3$Cu$_7$H$_4$ layers (@ degauss 0.02).

| System: □ | (Mg$_3$Cu$_7$H$_4$) | □ | λ | N(Ef) [states/Ry] | T$_C$ [K] |
|---|---|---|---|
| 10% electron doping | 0.7496 | 14.5960 | 10.052 |
| 0% doping | 0.7427 | 13.2089 | 11.577 |
| 10% hole doping | 0.7990 | 12.5702 | 10.399 |

### 6. References

1  J. P. Perdew, K. Burke and M. Ernzerhof, Generalized gradient approximation made simple, *Phys. Rev. Lett.*, 1996, **77**, 3865–3868.
2  J. P. Perdew, K. Burke and M. Ernzerhof, Generalized Gradient Approximation Made Simple [Phys. Rev. Lett. 77, 3865 (1996)], *Phys. Rev. Lett.*, 1997, **78**, 1396.
3  G. I. Csonka, J. P. Perdew, A. Ruzsinszky, P. H. T. Philipsen, S. Lebègue, J. Paier, O. A. Vydrov and J. G. Ángyán, Assessing the performance of recent density functionals for bulk solids, *Phys. Rev. B - Condens. Matter Mater. Phys.*, 2009, **79**, 155107.
4  P. E. Blöchl, Projector augmented-wave method, *Phys. Rev. B*, 1994, **50**, 17953–17979.
5  G. Kresse and D. Joubert, From ultrasoft pseudopotentials to the projector augmented-wave method, *Phys. Rev. B*, 1999, **59**, 1758–1775.
6  Vienna Ab initio Simulation Package, https://www.vasp.at.
7  S. Grimme, J. Antony, S. Ehrlich and H. Krieg, A consistent and accurate ab initio parametrization of density functional dispersion correction (DFT-D) for the 94 elements H-Pu, *J. Chem. Phys.*, 2010, **132**, 154104.
8  V. I. Anisimov, J. Zaanen and O. K. Andersen, Band theory and Mott insulators: Hubbard U instead of Stoner I, *Phys. Rev. B*, 1991, **44**, 943–954.
9  E. R. Ylvisaker, W. E. Pickett and K. Koepernik, Anisotropy and magnetism in the LSDA+U method, *Phys. Rev. B - Condens. Matter Mater. Phys.*, 2009, **79**, 35103.
10  B. Himmetoglu, A. Floris, S. De Gironcoli and M. Cococcioni, Hubbard-corrected DFT energy functionals: The LDA+U description of correlated systems, *Int. J. Quantum Chem.*, 2014, **114**, 14–49.
11  P. Giannozzi, S. Baroni, N. Bonini, M. Calandra, R. Car, C. Cavazzoni, D. Ceresoli, G. L. Chiarotti, M. Cococcioni, I. Dabo, A. Dal Corso, S. De Gironcoli, S. Fabris, G. Fratesi, R. Gebauer, U. Gerstmann, C. Gougoussis, A. Kokalj, M. Lazzeri, L. Martin-Samos, N. Marzari, F. Mauri, R. Mazzarello, S. Paolini, A. Pasquarello, L. Paulatto, C. Sbraccia, S. Scandolo, G. Sclauzero, A. P. Seitsonen, A. Smogunov, P. Umari and R. M. Wentzcovitch, QUANTUM ESPRESSO: A modular and open-source software project for quantum simulations of materials, *J. Phys. Condens. Matter*, 2009, **21**, 395502.
12  P. Giannozzi, O. Andreussi, T. Brumme, O. Bunau, M. Buongiorno Nardelli, M. Calandra, R. Car, C. Cavazzoni, D. Ceresoli, M. Cococcioni, N. Colonna, I. Carnimeo, A. Dal Corso, S. De Gironcoli, P. Delugas, R. A. Distasio, A. Ferretti, A. Floris, G. Fratesi, G. Fugallo, R. Gebauer, U. Gerstmann, F. Giustino, T. Gorni, J. Jia, M. Kawamura, H. Y. Ko, A. Kokalj, E. Kücükbenli, M. Lazzeri, M. Marsili, N. Marzari, F. Mauri, N. L. Nguyen, H. V. Nguyen, A. Otero-De-La-Roza, L. Paulatto, S. Poncé, D. Rocca, R. Sabatini, B. Santra, M. Schlipf, A. P. Seitsonen, A. Smogunov, I. Timrov, T. Thonhauser, P. Umari, N. Vast, X. Wu and S. Baroni, Advanced capabilities for materials modelling with Quantum ESPRESSO, *J. Phys. Condens. Matter*, 2017, **29**, 465901.
13  D. R. Hamann, Optimized norm-conserving Vanderbilt pseudopotentials, *Phys. Rev. B - Condens. Matter Mater. Phys.*, 2013, **88**, 85117.
14  M. Methfessel and A. T. Paxton, High-precision sampling for Brillouin-zone integration in metals, *Phys. Rev. B*, 1989, **40**, 3616–3621.
15  S. Baroni, S. De Gironcoli, A. Dal Corso and P. Giannozzi, Phonons and related crystal properties from density-functional perturbation theory, *Rev. Mod. Phys.*, 2001, **73**, 515–562.
16  P. B. Allen and R. C. Dynes, Transition temperature of strong-coupled superconductors reanalyzed, *Phys. Rev. B*, 1975, **12**, 905–922.
17  K. Momma and F. Izumi, VESTA: A three-dimensional visualization system for electronic and structural analysis, *J. Appl. Crystallogr.*, 2008, **41**, 653–658.
18  H. T. Stokes and D. M. Hatch, FINDSYM: Program for identifying the space-group symmetry of a crystal, *J. Appl. Crystallogr.*, 2005, **38**, 237–238.